\documentclass[useAMS,usenatbib]{mn2e}

\usepackage{lscape}
\usepackage{longtable}
\usepackage{graphicx}
\usepackage{siunitx}
\usepackage{url}
\usepackage{gensymb}
\usepackage{multirow}
\usepackage{amsmath}
\usepackage{breqn}
\usepackage{pdflscape}
\usepackage{booktabs}
\usepackage[table]{xcolor}
\usepackage{float}
\usepackage{caption}

\newcommand{\arp}{Arp 220}
\newcommand{\htwo}{H\textsc{ii}}

\newcommand{\emcee}{\textsc{emcee}}
\newcommand{\mn}{\textsc{multinest}}
\newcommand{\Lagr}{\mathcal{L}}
\newcommand{\Zagr}{\mathcal{Z}}
\newcommand{\iras}[1]{IRAS\,#1}

\newcommand{\affil}[1]{$^{\rm #1}$}
\newcounter{inst}
\newcommand{\inst}[1]{\noindent%
  \refstepcounter{inst}\affil{\arabic{inst}\label{#1}}     
  }

\begin{document}
\title[Modelling the radio emission of 19 starbursts]{The Spectral Energy Distribution of Powerful Starburst Galaxies I: Modelling the Radio Continuum}
\author[T. J. Galvin et al. 2015]{
T. J. Galvin\affil{\ref{WSU},\ref{CIRA},\ref{CASS}}, 
N. Seymour\affil{\ref{CIRA}}, 
J. Marvil\affil{\ref{CASS}},
M. D. Filipovi\'{c}\affil{\ref{WSU}}, 
N. F. H. Tothill\affil{\ref{WSU}}, \newauthor
R. M. McDermid\affil{\ref{MQ},\ref{AAO}},
N.~Hurley-Walker\affil{\ref{ICRAR}},
P.~J.~Hancock\affil{\ref{ICRAR},\ref{CAASTRO}},
J.~R.~Callingham\affil{\ref{USyd},\ref{CASS},\ref{CAASTRO}},\newauthor
R.~H.~Cook\affil{\ref{CIRA},\ref{UWA}},
R.~P.~Norris\affil{\ref{WSU},\ref{CASS}}
M.~E.~Bell\affil{\ref{CASS},\ref{CAASTRO}},
K.~S.~Dwarakanath\affil{\ref{RRI}},
B.~For\affil{\ref{UWA}}, \newauthor
B.~M.~Gaensler\affil{\ref{USyd},\ref{CAASTRO},\ref{Toronto}},
L.~Hindson\affil{\ref{VUW}},
M.~Johnston-Hollitt\affil{\ref{VUW},\ref{PSL}},
A.~D.~Kapi\'nska\affil{\ref{CAASTRO},\ref{UWA}},\newauthor
E.~Lenc\affil{\ref{USyd},\ref{CAASTRO}},
B.~McKinley\affil{\ref{CAASTRO},\ref{ANU}},
J.~Morgan\affil{\ref{ICRAR}},
A.~R.~Offringa\affil{\ref{CAASTRO},\ref{ASTRON}},
P.~Procopio\affil{\ref{CAASTRO},\ref{UMelb}},\newauthor
L.~Staveley-Smith\affil{\ref{CAASTRO},\ref{UWA}}, 
R.~B.~Wayth\affil{\ref{ICRAR},\ref{CAASTRO}},
C.~Wu\affil{\ref{UWA}},
Q.~Zheng\affil{\ref{VUW}}\\
{\small \inst{WSU}\,University of Western Sydney, Locked Bag 1797, Penrith, NSW, 2751, Australia}\\
{\small \inst{CIRA}\,International Centre for Radio Astronomy Research, Curtin
University, Bentley, WA 6102, Australia}\\
{\small \inst{CASS}\,CSIRO Astronomy and Space Science, Marsfield, NSW
1710, Australia}\\
{\small \inst{MQ}\,Department of Physics and Astronomy, Macquarie University, Sydney, NSW 2109, Australia}\\
{\small \inst{AAO}\,Australian Gemini Office, Australian Astronomical Observatory, PO Box 915, Sydney, NSW 1670, Australia}\\
{\small \inst{ICRAR}\,International Centre for Radio Astronomy Research,
Curtin University, Bentley, WA 6102, Australia}\\
{\small \inst{USyd}\,Sydney Institute for Astronomy, School of Physics,
The University of Sydney, NSW 2006, Australia}\\
{\small \inst{CAASTRO}\,ARC Centre of Excellence for All-sky Astrophysics
(CAASTRO)}\\
{\small \inst{VUW}\,School of Chemical \& Physical Sciences, Victoria
University of Wellington, Wellington 6140, New Zealand}\\
{\small \inst{UWA}\,International Centre for Radio Astronomy Research,
University of Western Australia, Crawley 6009, Australia}\\
{\small \inst{ASTRON}\,Netherlands Institute for Radio Astronomy (ASTRON),
PO Box 2, 7990 AA Dwingeloo, The Netherlands}\\
{\small \inst{RRI}\,Raman Research Institute, Bangalore 560080, India}\\
{\small \inst{Toronto}\,Dunlap Institute for Astronomy and Astrophysics,
50 St. George St, University of Toronto, ON M5S 3H4, Canada}\\
{\small \inst{PSL}\,Peripety Scientific Ltd., PO Box 11355 Manners Street, Wellington 6142, New Zealand}\\
{\small \inst{UMelb}\,School of Physics, The University of Melbourne,
Parkville, VIC 3010, Australia}\\
{\small \inst{ASU}\,School of Earth and Space Exploration, Arizona State
University, Tempe, AZ 85287, USA}\\
{\small \inst{ANU}\,Research School of Astronomy and Astrophysics,
Australian National University, Canberra, ACT 2611, Australia}\\
{\small \inst{Haystack}\,MIT Haystack Observatory, Westford, MA 01886, USA}\\
{\small \inst{CfA}\,Harvard-Smithsonian Center for Astrophysics, 60 Garden
Street, Cambridge, MA, 02138, USA}\\
{\small \inst{UDub}\,Department of Physics, University of Washington,
Seattle, WA 98195, USA}\\
{\small \inst{UWM}\,Department of Physics, University of
Wisconsin--Milwaukee, Milwaukee, WI 53201, USA}\\
{\small \inst{MIT}\,Kavli Institute for Astrophysics and Space Research,
Massachusetts Institute of Technology, Cambridge, MA 02139, USA}\\
{\small \inst{NCRA}\,National Centre for Radio Astrophysics, Tata
Institute for Fundamental Research, Pune 411007, India}\\
}

\pagerange{\pageref{firstpage}--\pageref{lastpage}} \pubyear{2017}

\maketitle

\begin{abstract}
We have acquired radio continuum data between 70\,MHz and 48\,GHz for a sample of 19 southern starburst galaxies at moderate redshifts ($0.067 < z < 0.227$) with the aim of separating synchrotron and free-free emission components. Using a Bayesian framework we find the radio continuum is rarely characterised well by a single power law, instead often exhibiting low frequency turnovers below 500\,MHz, steepening at mid-to-high frequencies, and a flattening at high frequencies where free-free emission begins to dominate over the synchrotron emission. These higher order curvature components may be attributed to free-free absorption across multiple regions of star formation with varying optical depths. The decomposed synchrotron and free-free emission components in our sample of galaxies form strong correlations with the total-infrared bolometric luminosities. Finally, we find that without accounting for free-free absorption with turnovers between 90 to 500\,MHz the radio-continuum at low frequency ($\nu < 200$\,MHz) could be overestimated by upwards of a factor of twelve if a simple power law extrapolation is used from higher frequencies. The mean synchrotron spectral index of our sample is constrained to be $\alpha=-1.06$, which is steeper then the canonical value of $-0.8$ for normal galaxies. We suggest this may be caused by an intrinsically steeper cosmic ray distribution. 
\end{abstract}

\section{Introduction}

Understanding the star formation history (SFH) of the Universe is one of the key science goals of the Square Kilometre Array \citep[SKA; ][]{2015aska.confE..67P} and its pathfinder projects \citep{2011PASA...28..215N}. radio continuum emission offers a unique advantage over other wavebands as it is impervious to the effects of dust attenuation and is able to provide an unbiased view into the star formation rates (SFR) of distant galaxies through cosmic time \citep{2008MNRAS.386.1695S,2015aska.confE..68J}. Current critical radio continuum SFR indicators, based mostly on the 1.4\,GHz luminosity, have been calibrated against far-infrared (FIR) measures using the far-infrared to radio correlation \citep[FRC; ][]{1992ARA&A..30..575C}. 

The FRC itself is a tight, linear relationship across many orders of magnitude between the far-infrared and radio continuum luminosities of star forming galaxies \citep{2001ApJ...554..803Y,2003ApJ...586..794B,2011ApJ...731...79M}. Its existence is comprised of three individual emission processes that are all manifestations of high mass star (HMS; M$_\odot > 8$\,M$_\odot$) formation.

FIR emission, spanning 40 to 500\,$\mu$m, originates from widespread dust cirrus heated by UV and optical emission from a combination of mostly young HMS and an older stellar population. The observed radio continuum is a superposition of two individual mechanisms, the most prominent at low frequencies being non-thermal synchrotron emission. This process is thought to be formed from the relativistic electrons, accelerated by the remnants of Type II and Type Ib supernova of HMS, gyrating within large scale galactic magnetic fields. Although synchrotron emission makes up roughly 90\% of the radio continuum at 1.4\,GHz of normal galaxies, it is a delayed tracer to SF, taking upwards of $10^7$ years for the electrons to diffuse \citep{1992ARA&A..30..575C}. 

Thermal free-free emission is the second mechanism that makes up the radio continuum. Its underlying process is powered by the ionisation of \htwo\ regions by UV flux from HMSs. Unlike synchrotron emission, it is a direct, near instantaneous tracer of SFR. Despite this, it is relatively unused as a radio continuum SFR indicator due to its flat spectral index ($\alpha=-0.1$, where $S\propto\nu^\alpha$) and the fact that at low frequencies, where the survey speeds of radio telescopes are most efficient, the spectrum is overwhelmingly dominated by synchrotron. Isolating it requires either model fitting using a broad, densely sampled radio continuum spectral energy distribution \citep[SED; ][]{1992ApJ...401...81P,2016MNRAS.461..825G}, or high frequency observations ($\nu > 20$\,GHz) where synchrotron emission is mostly absent \citep{2012ApJ...761...97M}. 

Although calibrating the radio continuum SFR measures through the FRC has proved effective in the local Universe ($z<0.15$), there remains considerable uncertainty as to how their reliability will scale with increasing redshifts. \citet{2009ApJ...706..482M} argues that due to a combination of the changing composition of the radio continuum with increasing frequencies and the suppression of synchrotron emission due to inverse-Compton scatter off the cosmic microwave background, scaling in proportion to $(1+z)^4$, that there should be an evolution in the observed frame FRC. Observational evidence presented by \citet{2010A&A...518L..31I} and \citet{2011ApJ...731...79M} both use image stacking techniques to demonstrate no change in the FRC up to redshifts of 2, suggesting that the physical origin of the FRC may be more complex than first throught. 

Future radio continuum surveys expected from the SKA and its pathfinder projects will explore the high redshift Universe ($z>1.0$). In this parameter space it is expected that distant, faint star forming galaxies (SFG), whose SFR are in excess of 100\,M$_\odot$\,yr$^{-1}$, will be the predominant class of object detected with these surveys. Work by \citet{2010MNRAS.405..887C} also shows that the effects of free-free absorption, particularly in the case of multiple star forming regions with different optical depths, will further complicate the observed radio continuum. Although synchrotron self absorption can produced turnover features that can complicated the observed SED, star forming galaxies do not have the required brightness temperatures \citep{1992ARA&A..30..575C}. Correctly interpreting the emission properties that trace star formation will require an improved understanding of the underlying physical mechanisms and how they can be characterised through their diverse SEDs. 

In this study, we investigate the intrinsic emission components of 19 powerful starforming luminous infrared galaxies (LIRGS), which are ideal representative sources of distant SFGs,  at redshifts between 0.0627 and 0.227. We construct a series of comprehensive radio continuum SEDs, ranging between 70\,MHz up to 48\,GHz in the observed frame with the aim of isolating the thermal free-free component and identifying the effects of free-free absorption (FFA) at low frequency. As the free-free emission is (1) a direct tracer of SFR, (2) exhibit a flat spectral slope, and (3) originates from the same \htwo\ regions as hydrogen recombination lines, it is an excellent candidate to craft SFR measures that are compatible with the high redshift SFGs that will be revealed with SKA and its pathfinder projects.  For our sample of objects we have also acquired optical spectroscopy data using the Wide-Field Spectrograph \citep[WiFeS; ][]{2007Ap&SS.310..255D,2010Ap&SS.327..245D}. We will analyse the optical spectroscopic data in conjunction with this radio continuum modelling in a subsequent series of papers. 

We assume a flat Universe, where $\Omega_m = 0.277$, $\Omega_\lambda = 0.733$ and \mbox{$H_0= 70.2 \mathrm{\ km\ s}^{-1} \mathrm{\ Mpc}^{-1} $} following \citet{2009ApJS..180..330K}.

\section{Data}
\subsection{Source Selection}

In this study we selected a sample of all known southern ($\delta< -30^\circ$) LIRGS, defined as having IR luminosities greater than L$_{8-1000\mu\mathrm{m}}>10^{11}\,\mathrm{L}_\odot$. These objects were specifically targeted due to their high SFR, as this implies that there would be a measurable thermal component in their radio continuum. These types of objects are analogous to the types of distant star forming galaxies which are expected to predominately comprise the next generation of future deep surveys.

The sample for this study was constructed using the Revised $IRAS$ Faint Source Catalog \citep{2014MNRAS.442.2739W}. We identified all sources with a $60\,\mu$m flux density in excess of 1.4\,Jy (S$_{60\,\mu \textrm{m}} >1.4\,$Jy) and a spectroscopic redshift in the range $0.067 < z < 0.227$. This was done, not only to target galaxies with high SFR, but to also allow for future ground based observations of the Paschen-$\alpha$ (Pa$\alpha$; $\lambda=1.875$\,$\mu$m) hydrogen recombination line, which is a relatively un-attenuated measure of star formation. Potential sources were cross referenced with the Sydney University Molonglo Sky Survey (SUMSS) catalog \citep{2003MNRAS.342.1117M,2013yCat.8081....0M} in order to obtain radio flux densities at 843 MHz. 

In order to construct a representative sample of SFGs, sources with a detectable AGN component, as seen in their optical spectra, or those flagged as a quasi-stellar objects by \citet{2014MNRAS.442.2739W}, were exclude from further consideration. We also scaled the SUMSS flux density measurement to 1.4\,GHz using the model described by \citet{1992ARA&A..30..575C} to assess potential AGN activity. Sources defined as radio or infrared excess (five times as much radio or infrared luminosities expected when considering the FRC) following \citet{2001ApJ...554..803Y} were also discarded, as such excess is indicative of the presence of AGN (see the $q$ parameter defined below). After excluding sources with some AGN indicator our final sample consisted of 20 sources. We list their positions, spectroscopic redshifts and IR luminosities in Table~\ref{Table:Sample}.

\begin{table}
\caption{The complete source sample used throughout this study. IRAS\ F14378$-$3651 was ultimately excluded from further processing due to LST constraints. \label{Table:Sample} }
\begin{tabular}{ccccc}
\toprule
  Name &
  RA &
  Dec &
  $z$ &
    L$_{8 - 1000\,\mu\mathrm{m}}$ \\
  {\em IRAS} &
  J2000   &
  J2000   &
               &
   Log\,L$_\odot$ \\

\midrule
  F00198$-$7926 & 00:21:53.6 & $-$79:10:07.79 & 0.07 & 12.12\\
  F00199$-$7426 & 00:22:07.0 & $-$74:09:41.89 & 0.10 & 12.22\\
  F01268$-$5436 & 01:28:47.7 & $-$54:21:25.62 & 0.09 & 11.97\\
  F01388$-$4618 & 01:40:55.9 & $-$46:02:53.32 & 0.09 & 12.08\\
  F01419$-$6826 & 01:43:17.1 & $-$68:11:24.12 & 0.08 & 11.8\\[0.1cm]
  
  F02364$-$4751 & 02:38:13.9 & $-$47:38:11.34 & 0.10 & 12.05\\
  F03068$-$5346 & 03:08:20.9 & $-$53:35:17.66 & 0.07 & 11.9\\
  F03481$-$4012 & 03:49:53.8 & $-$40:03:41.03 & 0.10 & 11.86\\
  F04063$-$3236 & 04:08:18.9 & $-$32:28:30.35 & 0.11 & 12.07\\
  F06021$-$4509 & 06:03:33.6 & $-$45:09:41.12 & 0.16 & 12.23\\[0.1cm]
  
  F06035$-$7102 & 06:02:54.1 & $-$71:03:10.48 & 0.08 & 12.15\\
  F06206$-$6315 & 06:21:01.2 & $-$63:17:23.81 & 0.09 & 12.2\\
  F14378$-$3651 & 14:40:59.0 & $-$37:04:32.24 & 0.07 & 12.07\\
  F18582$-$5558 & 19:02:24.0 & $-$55:54:08.56 & 0.07 & 11.63\\
  F20117$-$3249 & 20:14:55.3 & $-$32:40:00.50 & 0.10 & 11.92\\[0.1cm]
  
  F20445$-$6218 & 20:48:44.1 & $-$62:07:25.35 & 0.11 & 11.95\\
  F21178$-$6349 & 21:21:53.8 & $-$63:36:43.68 & 0.07 & 11.63\\
  F21292$-$4953 & 21:32:36.2 & $-$49:40:24.74 & 0.14 & 12.39\\
  F21295$-$4634 & 21:32:49.4 & $-$46:21:03.93 & 0.07 & 11.72\\
  F23389$-$6139 & 23:41:43.5 & $-$61:22:52.62 & 0.09 & 12.14\\
\bottomrule\end{tabular}

\end{table}

\subsection{ATCA Observations}
Over five non-consecutive nights, 19 of the 20 sources in our sample (IRAS\,F14378-3651 was dropped due to LST constraints) were observed across 11 central frequencies (Table~\ref{Table:CABB}) using the Australia Telescope Compact Array \citep[ATCA;][]{1992JEEEA..12..103F,2011MNRAS.416..832W} under the project code C2993 (PI: Galvin). With the Compact Array Broadband Backend (CABB) filters, a spectral window of 2.048\,GHz was available for each of the targeted central frequencies. In total this provided roughly 22.5\,GHz of coverage from 1.1 to 94.0\,GHz. We adopted a snapshot imaging approach due to the diverse LST range of our sample. To help optimise efficiency we grouped sources based on their positions to share phase reference calibrators. This was important as at high frequencies an increasingly large fraction of time is lost to calibration overheads (i.e. pointing calibrations and phase reference scans). 

\begin{table}
\centering
\caption{An overview of the ATCA data (Project code: C2993, PI: Galvin) obtained as part of this study. All data used the compact array broadband backend, giving a total of 2.048\,GHz per central frequency. We include the Largest Angular Scale (LAS) that each array is sensitive to. \label{Table:CABB}}
    \begin{tabular}{ c c c cc }
        \toprule
        Central Frequency & Band & Array & Date Observed & LAS\\ 
        GHz & & & & $``$\\
        \midrule
        2.1 & L/S  & 6A & 23-01-2015  & 89.0 \\ [0.1cm]
        5.0 & C/X & 6A & 27-01-2015  & 37.4 \\ 
        5.0 & C/X & 750C & 29-12-2015 & 275.3 \\ 
        6.8 & C/X & 6A & 27-01-2015 & 27.5 \\ 
        6.8 & C/X & 750C & 29-12-2015 & 201.7 \\ 
        8.8 & C/X & 6A & 27-01-2015 & 21.2 \\ 
        8.8 & C/X & 750C & 29-12-2015 & 155.9 \\ 
        10.8 & C/X & 6A & 27-01-2015 & 17.31 \\ 
        10.8 & C/X & 750C & 29-12-2015 &  127.0 \\ [0.1cm]
        17.0 & K & 6A & 23-01-2015 & 11.0 \\ 
        17.0 & K & 750C & 31-12-2015  & 80.9 \\
        17.0 & K & H168 & 06-09-2016 & 60.5 \\
        21.0 & K & 6A & 23-01-2015 & 8.9 \\ 
        21.0 & K & 750C & 31-12-2015 & 65.3 \\
        21.0 & K & H168 & 06-09-2016 & 49.0 \\ [0.1cm]
        45.0 & Q & H214 & 4-09-2014 & 17.0 \\ 
        47.0 & Q & H214 & 4-09-2014 & 15.9 \\ 
        89.0 & W & H214 & 4-09-2014 & 8.6 \\ 
        93.0 & W & H214 & 4-09-2014 & 8.2 \\ [0.1cm]
        \bottomrule
    \end{tabular}
\end{table}

The first night on the 4$^{\mathrm{th}}$ October 2014 targeted the Q and W band frequencies and was performed in a H214 hybrid configuration. This compact array configuration was selected to help prevent resolving out source structure. PKS\,1921$-$293 (RA,DEC J2000: 19:24:51.05, -29:14:30.12) was used as the bandpass calibrator while Uranus was used to provide a flux density scale. Due to the high frequency, pointing calibrations were performed between each slew greater then 10$^\circ$. A full hour angle synthesis was not possible due to the considerable overheads required with observing at these frequencies. Instead, we elected to observed each source for a single 15 minute exposure and measure the flux of each source in the $(u,v)$-plane exclusively. Although normally a single cut in the $(u,v)$-plane would introduce source confusion, the H214 hybrid array, with two antenna along the north-south spur, provided enough spatial coverage to sample the $(u,v)$-plane adequately enough to isolate our sources in the sky. Elevated path noise  \citep[a measure of the atmospheric phase stability; ][]{2006PASA...23..147M} only allowed us to observe 6 sources at the W band central frequencies. Ultimately these W band data were discarded due to difficulties during calibration.

Centimetre data were collected over a number of individual observing runs. Initially, L and K band data were collected on the 23$^{\mathrm{rd}}$ January 2015 in a 6A array configuration. We used PKS\,1934$-$638 to provide a flux density calibration for both bands. For K band data taken on this night, PKS\,1921$-$293 was used as a bandpass calibrator. During this initial 12 hour observing run each source was observed for at least 5 minutes across at least 3 cuts. A phase calibrator was also visited at least once every ten minutes. Subsequent K band data were collected on the 30$^{\mathrm{th}}$ December 2015 and 26$^{\mathrm{th}}$ September 2016 in compact 750C and H168 array configurations. Data obtained in the 750C array configuration used the same observing strategy outlined above. For the H168 array K band observing we elected to dwell on each source for a single 10 minute block of time across a single 4 hour block of unallocated telescope time. This `single block' approach minimised the total time lost to overheads while, with the addition of the north-south spur, adequately sampled the innermost $(u,v)$-region. A phase reference scan was made after each source.

The C/X band frequencies were obtained across two separate observing runs totalling roughly 17 hours. The first, performed on 27$^{\mathrm{th}}$ January 2015 for 12 hours, used a sparsely distributed 6A configuration. On 29$^{\mathrm{th}}$ December 2015 we again revisited the sample in a compact 750C array. For both observing sessions PKS 1934$-$638 was used as a bandpass calibrator and flux density scale. In total, across both observing runs each source was observed for roughly 7 minutes across  at least 4 cuts. 

Due to the wide range of LST of our sources we were not able to ensure a consistent amount of integration time equally spread across the $(u,v)$-space for our complete sample. Traditionally this would be a problem for image deconvolution due to the poorly constrained instrumental response, but as we are primarily interested in a known source at the phase centre of each pointing, this is not a critical issue. Hence our major obstacle is trying to prevent resolving source structure with increasing resolution. The inclusion of a short baseline data from 750C, H214 and H168 array configurations for the C/X and K bands and a natural weighting scheme helped in this regard. Collectively the combination of data were sensitive to roughly the same angular scales. 

\subsection{Murchison Widefield Array}

Low frequency data were obtained from the SKA-LOW precursor, the Murchison Widefield Array \citep[MWA; ][]{2009IEEEP..97.1497L,2013PASA...30....7T}. Located in Western Australia, it is comprised of 2048 dual polarisation dipole antennas capable of operating between 70 to 320\,MHz with an instantaneous frequency coverage of 30.72\,MHz. 

One of its key science products, entitled the GaLactic and Extragalactic MWA Survey \citep[GLEAM, ][]{2015PASA...32...25W}, is imaging the low frequency sky for declinations south of \texttt{+}30$^\circ$. The survey itself covers 30000 square degrees to a $\approx$90\% completeness level at 160\,mJy. A description of the observing, calibration, imaging and post-image calibration strategies to extract an extra-galactic source catalogue is presented by \citet{2017MNRAS.464.1146H}. GLEAM is the largest fractional bandwidth all-sky survey to date, with the final catalog containing twenty sub-band flux density measurements for each source across most of the MWA frequency range. The internal flux calibration is better than 3\% and is based on the \citet{1977A&A....61...99B} scale.

Source identification and extraction in GLEAM \citep[as outlined by ][]{2017MNRAS.464.1146H} was performed using the \textsc{aegean} software package \citep{2012MNRAS.422.1812H}. A deep image covering the frequency range of 170-231\,MHz was used to extract an initial reference catalog. After applying quality control measures, the flux density of each source in this reference catalog was then measured in the twenty 7.68\,MHz narrow-band images that span the 72 to 231\,MHz frequency range \citep{2017MNRAS.464.1146H}.

\subsubsection{Detected Sources}
\label{sec:mwa_detected}
We inspected the GLEAM catalogue to obtain possible low frequency flux densities for our sample of galaxies. Owing to the resolution of GLEAM, which is around 120$''$, we compared potential matches by eye\footnote{using the GLEAM postage stamp server found at \url{http://mwa-web.icrar.org/gleam_postage/q/form}} to ensure that they were genuine detections in a non-confused field. We found sources \iras{F00198$-$7926}, \iras{F01268$-$5436}, \iras{F02364$-$4751}, \iras{F03068$-$5346}, \iras{F03481$-$4012}, \iras{F04063$-$3236}, \iras{F21292$-$4953} and \iras{F23389$-$6139} had clear counterparts in the catalog. Source \iras{F06035$-$7102}, being in the direction of the Large Magellanic Cloud (LMC), was not included in this release of GLEAM. Subsequent source extraction for this source was carried out for a small region surrounding it using the GLEAM pipeline. Some sub-band measurements were discarded as they were described as having negative integrated flux densities. This was possible as the sensitive 170-231\,MHz reference image was used to identify sources, whose positions was fixed when the same sources were fitted in the noisier sub-band images.  

\subsubsection{Non-detections}
\label{sec:mwa_non-detected}
For sources in our sample without a reliable MWA detection, we used the \textsc{priorized} option available in \textsc{aegean} to estimate the flux density and uncertainty in the GLEAM broad band images. These broad band images were at central frequencies of 88, 115, 155, each with 30\,MHz of frequency coverage, and 200\,MHz with 60\,MHz of bandwidth. \textsc{priorized} allows the user to fix properties of some source (including its position) and specify characteristics of the MWA synthesised beam while fitting for an object. Using this method we were able to obtain a further set of low significance measurements for sources \iras{F00199$-$7426}, \iras{F01388$-$4618}, \iras{F01419$-$6826}, \iras{F06021$-$4509}, \iras{F06206$-$6315}, \iras{F18582$-$5558}, \iras{F20117$-$3249}, \iras{F20445$-$6218}, \iras{F21178$-$6349} and \iras{F21295$-$4634}. 

All MWA GLEAM measurements described in Sections~\ref{sec:mwa_detected} and \ref{sec:mwa_non-detected} are listed in Table~\ref{table:mwa_measurements}. 

\subsection{Archived radio continuum Data}

\begin{table}
\centering
	\caption{An overview of the archival ATCA data identified for this project. All observations used the pre-CABB ATCA correlator, providing only 128\,MHz of bandwidth. \label{atoa-data}}
	\begin{tabular}{ccccc}
	    \toprule
		Project & Date & $\nu$ & RA & DEC\\
		&& (GHz) &(J2000) & (J2000) \\
	    \midrule
	    \multirow{2}{*}{C222} & 21-6-1993 & 4.8 & 00:22:08.4& -74:08:31.95\\
	     & 21-6-1993 & 8.6 & 00:22:08.4& -74:08:31.95 \\
	    \midrule
	    \multirow{6}{*}{C539} & 5/7-1-1998 & 4.8 & 22:39:09.4 & -82:44:11.00 \\
	     & 5/7-1-1998  & 8.6 & 22:39:09.4 & -82:44:11.00\\
	     & 27/28-1-1998  & 4.8 & 22:39:09.4 & -82:44:11.00\\
	     & 27/28-1-1998  & 8.6 & 22:39:09.4 & -82:44:11.00\\ 
	     & 27-1-2002  & 1.4 & 22:39:09.4 & -82:44:11.00\\
	     & 27-1-2002  & 2.5 & 22:39:09.4 & -82:44:11.00\\   
	    \bottomrule
	   	    
	\end{tabular}
\end{table}

The Australian Telescope Online Archive\footnote{\url{http://atoa.atnf.csiro.au/}} (ATOA) was used to search for existing ATCA data of sources in our sample. Projects C222 and C593 were found to have observed IRAS\,F00199$-$7426 and IRAS\,F23389$-$6139 respectively. Their bandwidth was limited to 128\,MHz as they were taken using the pre-CABB ATCA correlator. We summarise these observations in Table~\ref{atoa-data}.

\subsection{Other data}

We sourced any radio continuum or far infrared flux density measurements from the literature for each of our sources. Initially, we collected measurements from the photometry tables for our sample tracked by the online NED\footnote{\url{http://ned.ipac.caltech.edu/}} tool.

Additional far infrared measurements were obtained from the \textit{AKARI} space telescope \citep{2007PASJ...59S.369M}. With the exception of \iras{F21295$-$4634} and \iras{F23389$-$6139} all sources in our sample were detected at 90\,$\mu$m in the \textit{AKARI} All-Sky Survey Point Source Catalog \citep{2010yCat.2298....0Y}. Brighter sources were also detected at 65 and 140\,$\mu$m. These measurements were not included in the photometry tables retrieved from NED. We list all measurements that we obtained from either NED, with references to their origin, or \citet{2010yCat.2298....0Y} in Table~\ref{table:ned_measurements}.

An image of the LMC at 20\,cm presented by \citet{2006MNRAS.370..363H,2007MNRAS.382..543H} was used to obtain a single flux density measurement at 1.4\,GHz for \iras{F06035$-$7102}. This was particularly important, as our L/S band ATCA data for this source was difficult to image, and ultimately discarded, due to its sparse $(u,v)$-sampling and the complexity of the LMC field. 

\section{Data Reduction}

\subsection{ATCA Radio Continuum}

The \textsc{miriad} \citep{1995ASPC...77..433S} and \textsc{karma} \citep{1996ASPC..101...80G} software packages were used for data reduction and analysis of the ATCA data. The guided automated flagging \textsc{miriad} routine \textsc{pgflag} was used in conjunction with more traditional \textsc{miriad} flagging and calibration tasks in order to perform an initial data reduction. Given the wide bandwidth of the CABB system, appropriate \textsc{miriad} tasks used the \textsc{nfbin} option to derive a frequency dependent calibration solution.

Once a calibration solution was applied to each of the observations program sources, the centimetre data were then imaged individually across all frequency bands using their complete bandwidth ($\Delta\nu~=~2.048$\,GHz, minus the edge channels automatically flagged by \textsc{atlod}). A Briggs robust parameter value of 2, corresponding to natural weighting, was used to provide the maximum signal-to-noise at the cost of producing a larger synthesised beam. Given the large fractional bandwidth provided by CABB, \textsc{mfclean} \citep{1994A&AS..108..585S} was used to deconvolve the multi-frequency synthesised dirty map. \textsc{miriad} tasks \textsc{restor} and \textsc{linmos} were used in conjunction to deconvolve sidelobe artefacts and performed primary beam correction while accounting for the spectral index of the clean components. These preliminary images were produced in order to inspect and compare the applied calibration solution among the $uv$-datasets for each source. 

\label{section:rp}

An iterative procedure, similar to that used by \citet{2016MNRAS.461..825G}, was used to exploit the generous 2\,GHz of bandwidth provided by the ATCA CABB system. Initially each CABB band was imaged individually using the recipe outlined above. Next, the \textsc{miriad} task \textsc{imfit} was used to constrain each source of interest using a single point source model. If the extracted peak flux density was above a signal to noise ratio (SNR) of 8 than the CABB dataset would be split into an increasing number of sub-bands and reprocessed. We also ensured that each sub-band had a fraction bandwidth larger then 10\% so that \textsc{mfclean} could safely be used. Given sufficient signal to noise across all sub-bands, this iterative procedure would continue to a maximum of four sub-bands. We utilised the \textsc{line} parameter in \textsc{invert} to ensure each image shared an equal amount of un-flagged channels. With such an approach, sources with high SNR were split into multiple data points, which could be used to better constrain the radio continuum emission models (see \S~\ref{section:modelling}). 

For high frequency Q and W band observations, we used the \textsc{miriad} task \textsc{uvfit} to fit a single point source model directly to the $(u,v)$-data for each source. We elected to not iteratively increase the number of sub-bands (similar to the process outlined above) or include a spectral index as a parameter while fitting to the visibilities (implemented in the \textsc{miriad} task \textsc{uvsfit}) as  at these frequencies, where the fractional bandwidth is below 5\% and spectral variation would be difficult to constrain. 

For archived data, where only 128\,MHz of data were available, the \textsc{nfbin} option was not used during typical calibration procedures\footnote{http://www.atnf.csiro.au/computing/software/miriad/userguide/}. A joint deconvolution method was applied to applicable datasets, namely those from C539, to minimise the resulting noise characteristics. Otherwise normal imaging procedures were used to deconvolve the beam response and apply primary beam corrections to all images. The task \textsc{imfit} was used to fit a point source model to sources of interest. Image residuals were inspected to ensure an adequate fit. 

\subsubsection{Resolving structure}

We examined the outputs of the iterative imaging process, including the modelled point source residuals, SUMSS images and the initial SEDs that the imaging pipeline produced to assess whether our data was resolving components of an object. This review showed that our 4\,cm data for \iras{F06035$-$7102} was detecting extended structure distinct from the main component of the source and within the SUMSS source. Therefore, we applied a convolving beam of $45\times45''$ (the same size of the SUMSS restoring beam) to all images above a frequency of 4\,GHz for this source, which was used to extract peak flux densities from. We added an additional 10\% error in quadrature for these measurements.

To assess whether there were other sources in our sample with similar diffuse features, we compared the peak fluxes obtained by fitting a point source model to all images before and after they were convolved with a $45\times45''$ Gaussian kernel, as well as the integrated flux of a Gaussian model fitted to the non-convolved image. We found that there was weak evidence of structure for \iras{F21292-4953} above frequencies of 6\,GHz. Convolved peak flux density measurements were therefore used for images between $4.0 < \nu < 22.0$ for this source. Otherwise, there were no other sources showing flux densities that were inconsistent among these methods. 

\iras{F23389$-$6139}, however, showed that the 4\,cm C/X bands was roughly $\sim6$\,mJy below the ATCA pre-CABB fluxes from project C539 and the trend seen between 3 to 17\,GHz. When investigating, we found that measurements made using the visibilities directly with the \textsc{miriad} task \textsc{uvfit} produced results that were in excellent agreement to the rest of the data. We believe that this difference in peak flux densities was the combination of clean bias and imaging artefacts that could not be deconvolved due to the sparse $(u,v)$-sampling. 

\label{sec:bt}
Our typical restoring beams were $20\times10''$ in L-band, $10\times5''$ in the C/X band and $5\times3''$ in the K-band. For each image we also computed the brightness temperature sensitivity. We compared this to the model from \citet{1992ARA&A..30..575C} normalised to 1\,K at 1.4\,GHz, the median brightness temperature of a face on spiral galaxy \citep{1998AJ....115.1693C}. The brightness temperatures of our images were all higher then this lower limit.

All ATCA flux density measurements obtained under the project code C2993 are listed in Table~\ref{table:atca_measurements}.

\section{SED Modelling}
\subsection{Variability}
Multi-epoch observations and source variability could give a false impression of curvature or complexity in an observed SED. For this study the majority of our data were collected within a two year timespan. SFGs do not show variability on such timescales at our sensitivities \citep{2016ApJ...818..105M}. MWA GLEAM Data Release One (DR1) conducted its observing campaign between August 2013 to July 2014. Over this timeframe multiple drift scans were performed across the southern sky before combining all available data into the final image set. Likewise, the majority of our ATCA data were taken between September 2014 and February 2015, with selected frequency bands being observed in compact array configurations up to September 2016. 

\subsection{Radio Continuum Models}
\label{section:modelling}
Given the broad coverage of our radio continuum data, which covers 70\,MHz to 48\,GHz in the observed frame, and size of our sample we elected to fit a series of increasingly complex models to all sources. All modelling was performed in the rest frame with a reference frequency, unless stated otherwise, of $\nu_0=1.4$\,GHz. 

\subsubsection{Power Law}
Initially, we fit a simple power law (which we label as `PL' in subsequent tables and figures) to all available flux density measurements, in the form of: 

\begin{equation}
\label{eq:pl}
	S_\nu = A\left(\frac{\nu}{\nu_0}\right)^\alpha.
\end{equation}

The terms $A$ and the spectral index, $\alpha$, are treated as free parameters and represent a normalisation component and the gradient in logarithmic space.

\subsubsection{Synchrotron and Free-Free Emission}

We can model the radio continuum as the sum of two distinct power laws. One representing the steep spectrum non-thermal synchrotron emission, and the second describing the flat spectral thermal free-free emission, following the form:

\begin{equation}
S_\nu = A\left(\frac{\nu}{\nu_0}\right)^{\alpha} + B\left(\frac{\nu}{\nu_0}\right)^{-0.1}, 
\label{eq:synandff}
\end{equation}

\noindent where $A$ and $B$ are treated as free parameters and represent the synchrotron and free-free normalisation components respectively. The free parameter $\alpha$ represents the synchrotron spectral index and, depending on the history of injected cosmic rays, is known to vary \citep{1997A&A...322...19N}. This model describes both synchrotron and free-free emission components as being completely optically thin (i.e. no curvature at low frequencies).  We label this model as `SFG NC'. 

\subsubsection{Synchrotron and Free-Free Emission with Free-Free Absorption}

When synchrotron and free-free emission are in a coextensive environment, synchrotron emission can be attenuated by free-free absorption (FFA) processes producing a low frequency turnover. This attenuation is influenced by the flux density, density and spatial distribution of the ionised free-free emission with respect to the non-thermal synchrotron emission. If the frequency of this turnover from free-free absorption is parameterised by $\nu_{t,1}$, then the optical depth can be described as $\tau=\left(\nu/\nu_{t,1}\right)^{-2.1}$. Following \citet{1992ARA&A..30..575C} and \citet{2010MNRAS.405..887C}, we describe this more complete model (labeled as `C' throughout) as:

\begin{equation}
\begin{split}
S_\nu = \left(1-e^{-\tau}\right)\left[B+A	\left(\frac{\nu}{\nu_{t,1}}\right)^{0.1+\alpha}\right]\left(\frac{\nu}{\nu_{t,1}}\right)^2,
\label{model4}
\end{split}
\end{equation}

\noindent where $\nu_t$ is the turn-over frequency where the optical depth reaches unity and $\alpha$ is the spectral index of the synchrotron emission. $A$ and $B$ represent the synchrotron and free-free emission components. We fit for $A, B, \nu_{t,1}$ and $\alpha$ simultaneously. To minimise model degeneracy, particularly in the case when normalisation components are subject to the $\nu^2$ scaling in the optically thick regime, we replace the $\nu_{0}$ term, set to 1.4\,GHz in other models, to instead be the turnover frequency parameter for each component.  

\subsubsection{Multiple Free-Free absorption components}

Model `C' assumes a single volume of thermal free-free plasma intermixed with synchrotron emission produced by relativistic electrons. Although this model was derived from observations of the irregular, clumpy galaxy Markarian 325 \citep{1990ApJ...357...97C}, \citet{2010MNRAS.405..887C} present a set of luminous infrared galaxies whose radio continuum show a number of high frequency `kinks' which could be attributed to multiple turnover features. Their interpretation suggests that when multiple star forming regions with different compositions or geometric orientations are integrated over by a large synthesised beam, such is the case of an unresolved galaxy, the observed radio continuum could be complex.   

Following this, we include an additional set of increasingly complex models that aim to capture these higher order features. 

First, we assume a single relativistic electron population that produces the synchrotron emission, that is inhomogeneously mixed with two distinct regions of star formation with distinct optical depths. This model (labeled `C2 SA') may be described as:

\begin{equation}
\begin{split}
	S_\nu = \left(1-e^{-\tau_1}\right)\left[B+A	\left(\frac{\nu}{\nu_{t,1}}\right)^{0.1+\alpha}\right]\left(\frac{\nu}{\nu_{t,1}}\right)^2 + \\
	\left(1-e^{-\tau_2}\right)\left[D+C	\left(\frac{\nu}{\nu_{t,2}}\right)^{0.1+\alpha}\right]\left(\frac{\nu}{\nu_{t,2}}\right)^2,
\end{split}
\end{equation}

\noindent where $\tau_1$ and $\tau_2$ describe the optical depths of component one and two (each parameterised with their own turnover frequency $\nu_{t,1}$ and $\nu_{t,2}$), $A$ and $C$ are the normalisation parameters for the synchrotron mechanism, and $B$ and $D$ scale the free-free component. $\alpha$ is the spectral index of the single synchrotron population. 

To account for sources where the low frequency SED does not indicate a turnover due to free-free absorption, we construct a model similar to `C2 1SA' in the form of:

\begin{equation}
\begin{split}
	S_\nu = \left(\frac{\nu}{\nu_0}\right)^{-2.1}\left[B+A	\left(\frac{\nu}{\nu_0}\right)^{0.1+\alpha}\right]\left(\frac{\nu}{\nu_0}\right)^2 + \\
	\left(1-e^{-\tau_2}\right)\left[D+C	\left(\frac{\nu}{\nu_{t,2}}\right)^{0.1+\alpha}\right]\left(\frac{\nu}{\nu_{t,2}}\right)^2.
\end{split}
\end{equation}

The model and its parameters, with the exception of $\tau_1$ which has been removed, behave the same way as `C2 1SA'.  The reference frequency for the low frequency component is parametrised as $\nu_0$ and set to 1.4\,GHz. We maintain this form as it allows the parameters $A$ and $B$ to be more directly comparable to $C$ and $D$. We label this model as `C2 1SAN'.

Next, we relax the single spectral index constraint. Although this introduces an additional parameter, its physical motivation is based on a galaxy merger, where two distinct systems merging drives a new burst of star formation. The electron distribution could, in such a scenario, be comprised of two different populations. We express this model as:

\begin{equation}
\label{eq:c2}
\begin{split}
	S_\nu = \left(1-e^{-\tau_1}\right)\left[B+A	\left(\frac{\nu}{\nu_{t,1}}\right)^{0.1+\alpha}\right]\left(\frac{\nu}{\nu_{t,1}}\right)^2 + \\
	\left(1-e^{-\tau_2}\right)\left[D+C	\left(\frac{\nu}{\nu_{t,2}}\right)^{0.1+\alpha_2}\right]\left(\frac{\nu}{\nu_{t,2}}\right)^2,
\end{split}
\end{equation}

\noindent where parameters carry the same meaning as in `C2 SA' except we introduce parameters $\alpha$ and $\alpha_2$ to characterise the synchrotron spectral indices of component one and component two respectively. We label this model simply as `C2'.

\subsection{Far Infrared Emission}

For normal type galaxies heated dust, traced by far infrared emission and approximated well by a greybody, begins to contribute a non-neglible fraction of the observed continuum at frequencies above 100\,GHz \citep{1992ARA&A..30..575C}. A greybody is an optically thin, modified blackbody spectrum written as: 

\begin{equation} 
\label{eq:gbmodel}
S_\nu\left(\lambda\right) =I \times \left[\left( \frac{60\,\mu\mathrm{m}}{ \lambda}\right)^{3+\beta}\times\frac{1}{e^{\frac{hc}{\lambda kT}}-1} \right],
\end{equation}

\noindent where $S_\nu$ is the flux density in Jy at frequency $\nu$, $T$ is the absolute temperature of the body in Kelvin, $\beta$ represents the power-law variation of the emissivity with wavelength, and $I$ is a normalisation. The $\beta$ component encodes properties of the distribution of dust grains and their sizes  with typical values in the range of 1 to 2 \citep{1983QJRAS..24..267H,2013MNRAS.436.2435S}. To appropriately constrain these additional free parameters we collect all measurements for each source up to $\lambda=500\,\mu\mathrm{m}$ available from the literature (summarised in Table~\ref{table:ned_measurements}). 

For each radio continuum model described in equations \ref{eq:pl}--\ref{eq:c2} we add a greybody component. The near orthogonal free-free and infrared emission components (specifically the Rayleigh-Jeans property of the greybody) allows us to reduce associated uncertainties for the thermal free-free emission while fitting each SED. 

\subsection{Fitting and Selection}
\subsubsection{Model Fitting}
While fitting our SED models, we followed a similar fitting approach as described by {\citet{2015ApJ...809..168C}}. An `affine invariant' Markov chain Monte Carlo ensemble sampler \citep{GW}, implemented as the \emcee\footnote{\url{https://github.com/dfm/emcee}}\ \textsc{python} package \citep{2013PASP..125..306F}, was used to constrain each of described radio continuum models for each source in our sample. This particular sampling method offers an efficient method of searching a parameter space, using a set of `walkers', for regions of high likelihood during model optimisation. These walkers are relatively insensitive to dependancies or covariance among the free parameters being optimised. The samples these walkers draw from some parameter space can be be marginalised over to estimate the probability density function of a set of parameters. In the Bayesian sense this sampled space is referred to as the posterior distribution. 

Assuming independent measurements whose errors are normally distributed, the log likelihood function that \emcee\ attempts to maximise is expressed as:

\begin{equation}
	\mathrm{ln}\,\Lagr\left(\theta\right) = -\frac{1}{2}\sum_{n} \left[\frac{\left(D_n-f\left(\theta\right)\right)^2}{\sigma_n^2} + \mathrm{ln}\left(2\pi\sigma_n^2\right)\right],
	\label{eq:like}
\end{equation} 

\noindent where $D$ and $\sigma$ are two vectors of length $n$ containing a set of flux density measurements and their associated uncertainties, and $f\left(\theta\right)$ is the model to optimise using the parameter vector $\theta$. 

As stated in Section~5.4 of \citet{2017MNRAS.464.1146H}, MWA GLEAM 7.68\,MHz sub-band measurements have correlated errors, which violates an underlying assumption of Equation~\ref{eq:like}. This covariance was introduced by a combination of their methods of applying primary beam, absolute flux scaling and ionosphere corrections and self-calibrating visibility data across 30.72\,MHz before creating the final set of twenty sub-band images with 7.68\,MHz widths. Some of these corrections have a direction dependent component, meaning the degree of correlation among sub-bands can vary as a function of position. Although this could be accounted for with an appropriate covariance matrix, whose off-diagonal elements represent the degree of correlation for a pair of sub-band fluxes, at present such a matrix is not known. Without accounting for this, any inferences made from constrained models could be biased or incorrect. 

We therefore adopted as part of our fitting routines a Mat\'ern covariance function \citep{gaussianprocess}, which aims to model the off diagonal elements of the unknown MWA GLEAM data covariance matrix. This is a radial type covariance function which assumes that measurements closer together (for our problem closer together in frequency space) are more correlated that those further apart. The form we adopt while performing all SED fitting is:

\begin{equation}
	k\left(r\right) = a^2\left(1+\frac{\sqrt{3}r}{\gamma} \right)\mathrm{exp}\left(-\frac{\sqrt{3}r}{\gamma}\right),
	\label{eq:matern}
\end{equation}

\noindent where $k$ is the parameterised Mat\'ern covariance function, $r$ is the $\Delta\nu$ between a pair of flux density measurements. $a$ and $\gamma$ are quantities constrained by \emcee. The \textsc{python} module \textsc{george}\footnote{\url{https://github.com/dfm/george}} \citep{hodlr} was used to implement and manage the Mat\'ern covariance function and supply the log likelihood,  for only the GLEAM flux density measurements, of some model given $\theta$. This was then summed with the log likelihood obtained using Equation~\ref{eq:like} for the independent flux density measurements and $\theta$ parameter vector. Note that the addition of $a$ and $\gamma$ increased the free parameters for each model by two. This covariance matrix modelling was not used for sources with a single MWA GLEAM flux density measurement. 

\subsubsection{Model Priors}

When constraining models within a Bayesian framework,  `priors' describe any known or likely conditions for each parameter in some $\theta$ set. Such priors can be as simple as limits to enforce a strict value range, or as complex as defining some distribution that the `true' value of a parameter is likely to take. The sampled posterior that the walkers construct can be sensitive to the conditions encoded as parameter priors, particularly if complex prior distributions are used. Therefore we use uniform priors which simply enforce a range of values some parameter is allowed to take. Uniform priors are also referred to as being `uninformed' as no likely distribution has been suppled to the Bayesian fitting frameworks.

Throughout our model fitting we ensure that normalisation parameters A, B, C and D remain positive, that the spectral index parameters $\alpha$ and $\alpha_2$ remain in the range of $-0.2>\alpha>-1.8$,  the turnover frequencies are between 10\,MHz to 40\,GHz, and the $a$ and $\gamma$ parameters of Equation~\ref{eq:matern} are between $-500$\,mJy to 500\,mJy and 1 to 200\,MHz respectively.

These priors are founded on the sound assumptions that flux densities are positive emission processes and we can only constrain turnovers within the range where we have data (note that some SED begins to flatten before the optical depth reaches unity). We construct the limits of the spectral index parameters $\alpha$ and $\alpha_2$ to allow a diverse ranges of values in the literature \citep{1990ApJ...357...97C,1997A&A...322...19N,2010MNRAS.405..887C}. For the Mat\'ern covariance parameters $a$ and $\gamma$ we make no assumption about their value and set their priors broad enough such that to encompass all GLEAM data.

\subsubsection{Model Selection}

\tabcolsep=0.15cm
\begin{table}
\caption{An overview of the natural log of the Bayes odds ratio from the \mn\ fitting of each model to every source. For each source, the values presented below are the evidence values for each model divided by the most preferred model (i.e. model with highest evidence value). As the natural log of ratio is presented, the most preferred models have values in this table equal to log$_e\left(1\right)=0$ (Bold-italic typeface with blue background). Less preferred models therefore have more negative numbers. Models where the ratio is less than log$_e\left(3\right)=1.1$ are considered indistinguishable from the most preferred model (italic typeface with green background).\label{Table:evidence}}

\begin{tabular}{cllllll}
\toprule
Source & PL & SFG & C & C2 & C2 & C2  \\
 {\em IRAS} &   & NC &   & 1SAN & 1SA &  \\
\midrule
F00198-7926 & -12.0 & -13.2 & -15.8 & -10.4 & \cellcolor{blue!25}\textit{\textbf{0.0}} & \cellcolor{green!25}\textit{-0.9}\\
F00199-7426 & -15.4 & -17.0 & -1.4 & \cellcolor{blue!25}\textit{\textbf{0.0}} & -1.9 & -2.2\\
F01268-5436 & -7.1 & \cellcolor{blue!25}\textit{\textbf{0.0}} & -2.2 & -4.7 & -5.1 & -5.2\\
F01388-4618 & -12.6 & -14.3 & \cellcolor{blue!25}\textit{\textbf{0.0}} & -3.0 & -3.7 & -4.5\\
F01419-6826 & \cellcolor{blue!25}\textit{\textbf{0.0}} & -1.4 & \cellcolor{green!25}\textit{-0.9} & -2.3 & -2.9 & -2.0\\
F02364-4751 & -28.5 & -30.1 & \cellcolor{blue!25}\textit{\textbf{0.0}} & -2.6 & -2.4 & -1.9\\
F03068-5346 & -9.4 & \cellcolor{green!25}\textit{-0.3} & \cellcolor{blue!25}\textit{\textbf{0.0}} & \cellcolor{green!25}\textit{-0.9} & -2.6 & -1.6\\
F03481-4012 & \cellcolor{blue!25}\textit{\textbf{0.0}} & -1.4 & -3.5 & -6.5 & -3.8 & -4.5\\
F04063-3236 & -23.2 & -24.8 & -30.4 & -9.5 & \cellcolor{blue!25}\textit{\textbf{0.0}} & -1.3\\
F06021-4509 & -8.5 & -10.2 & -10.4 & -1.8 & \cellcolor{green!25}\textit{-0.0} & \cellcolor{blue!25}\textit{\textbf{0.0}}\\
F06035-7102 & -53.9 & -55.1 & -21.7 & \cellcolor{blue!25}\textit{\textbf{0.0}} & -1.7 & -5.4\\
F06206-6315 & -65.8 & -67.2 & -29.4 & -23.1 & \cellcolor{blue!25}\textit{\textbf{0.0}} & \cellcolor{green!25}\textit{-0.3}\\
F18582-5558 & -24.9 & -26.5 & -22.9 & \cellcolor{green!25}\textit{-0.3} & \cellcolor{green!25}\textit{-1.1} & \cellcolor{blue!25}\textit{\textbf{0.0}}\\
F20117-3249 & -148.5 & -150.0 & -18.7 & \cellcolor{blue!25}\textit{\textbf{0.0}} & -7.5 & -4.3\\
F20445-6218 & -1.9 & -3.3 & \cellcolor{blue!25}\textit{\textbf{0.0}} & -1.3 & -3.3 & -1.3\\
F21178-6349 & -2.5 & -3.2 & \cellcolor{blue!25}\textit{\textbf{0.0}} & -4.1 & -1.6 & -2.1\\
F21292-4953 & \cellcolor{blue!25}\textit{\textbf{0.0}} & -1.8 & -4.3 & -3.3 & -5.9 & -5.9\\
F21295-4634 & -20.3 & -21.9 & \cellcolor{green!25}\textit{-0.1} & \cellcolor{blue!25}\textit{\textbf{0.0}} & -2.0 & \cellcolor{green!25}\textit{-1.0}\\
F23389-6139 & -1451.1 & -1452.6 & -211.6 & -146.5 & \cellcolor{blue!25}\textit{\textbf{0.0}} & \cellcolor{green!25}\textit{-1.1}\\
\bottomrule
\end{tabular}	
\end{table}

A Bayesian framework grants the ability to objectively test whether the introduction of additional model complexity (where additional complexity isn't restricted to an increasing set of nested models) is justified by an improved fit that isn't simply a symptom of overfitting. The evidence value, $\Zagr$, is defined as the integral of the complete parameter space. Although computationally difficult to numerically compute, especially in the case of increasing parameter dimensions, recent algorithms have proven to be adept at obtaining reliable estimates of its value. \mn\ \citep{2009MNRAS.398.1601F} uses a nested sampling method to obtain an estimate of the $\Zagr$ value. 

Given the $\Zagr$ values of competing models $M_1$ and $M_2$, one is able to determine whether a model is preferred over another given a set of data. The Bayes odds ratio between the evidence values $\Zagr_1$ and $\Zagr_2$ for models $M_1$ and $M_2$ is constructed as:

\begin{equation}
	\Delta\Zagr = e^{\left(\Zagr_1-\Zagr_2\right)}.
\end{equation}

The evidence supporting $M_1$ over $M_2$ is considered `very strong' with a $\Delta\Zagr$ in excess of 150. If $\Delta\Zagr$ is between $150>\Delta\Zagr >20$ or $20>\Delta\Zagr>3$ then it is seen as either `strong' or `positive' evidence (respectively) supporting $M_1$ over $M_2$. When $\Delta\Zagr$ is less than 3, then $M_1$ and $M_2$ are indistinguishable from one another. This scale was established by \citet{kass1995bayes} and is considered the standard ladder for preferred model selection.

We summarise the results of the Bayes odds ratio test for all models in Table~\ref{Table:evidence} and highlight the most preferred model with any of its competitors. While estimating $\Zagr$ for each model, \mn\ was configured to search the same parameter space as \emcee.

\section{Results}
\subsection{Model Results}

\tabcolsep=0.1cm
\begin{table*}
\centering
\caption{An overview of the most preferred models judged strictly by their evidence value and their constrained values. We use the 50$^{\mathrm{th}}$ percentile of the samples posterior distribution as the nominal value, and use the 16$^{\mathrm{th}}$ and 84$^{\mathrm{th}}$ percentiles to provide the $1\sigma$ uncertainties. Parameters not included in a model are marked by a `--'. We omit parameters constrained that belong to the Mat\'ern covariance function.  \label{Table:param}}
\begin{tabular}{cllllllllllll}
\toprule
Source & Model & A & B & $\alpha$ & $\nu_{t,1}$ & C & D & $\alpha_2$ & $\nu_{t,2}$ & I & Temp. & $\beta$  \\
{\em IRAS} &  & mJy & mJy &  & GHz & mJy & mJy &  & GHz & Jy & K &  \\
\midrule
F00198$-$7926 & C2 1SA & $187.9^{+16.7}_{-15.2}$ & $0.5^{+0.6}_{-0.4}$ & $-1.3^{+0.1}_{-0.1}$ & $0.2^{+0.0}_{-0.0}$ & $7.6^{+0.8}_{-1.0}$ & $0.5^{+0.4}_{-0.3}$ & - & $6.2^{+0.7}_{-0.6}$ & $0.23^{+0.06}_{-0.06}$ & $55.5^{+4.5}_{-3.2}$ & $1.3^{+0.3}_{-0.2}$\\
F00199$-$7426 & C2 1SAN & $6.4^{+2.6}_{-3.0}$ & $0.2^{+0.3}_{-0.2}$ & $-0.8^{+0.0}_{-0.0}$ & - & $46.6^{+21.8}_{-13.7}$ & $0.2^{+0.3}_{-0.2}$ & - & $0.5^{+0.2}_{-0.1}$ & $1.52^{+0.41}_{-0.29}$ & $40.1^{+1.8}_{-1.9}$ & $1.1^{+0.2}_{-0.1}$\\
F01268$-$5436 & SFG NC & $0.8^{+0.2}_{-0.2}$ & $12.4^{+0.4}_{-0.4}$ & $-1.0^{+0.0}_{-0.0}$ & - & - & - & - & - & $0.34^{+0.22}_{-0.12}$ & $44.7^{+4.6}_{-4.4}$ & $1.3^{+0.4}_{-0.3}$\\
F01388$-$4618 & C & $44.4^{+5.5}_{-4.2}$ & $0.2^{+0.2}_{-0.1}$ & $-0.7^{+0.0}_{-0.0}$ & $0.3^{+0.0}_{-0.0}$ & - & - & - & - & $0.59^{+0.25}_{-0.19}$ & $45.7^{+3.8}_{-3.1}$ & $1.6^{+0.2}_{-0.3}$\\
F01419$-$6826 & PL & $8.7^{+0.3}_{-0.3}$ & - & $-0.7^{+0.0}_{-0.0}$ & - & - & - & - & - & $0.82^{+0.37}_{-0.32}$ & $41.2^{+3.7}_{-2.5}$ & $1.7^{+0.2}_{-0.4}$\\
F02364$-$4751 & C & $86.5^{+4.7}_{-4.6}$ & $0.3^{+0.3}_{-0.2}$ & $-0.8^{+0.0}_{-0.0}$ & $0.3^{+0.0}_{-0.0}$ & - & - & - & - & $1.04^{+0.43}_{-0.30}$ & $40.8^{+2.9}_{-2.4}$ & $1.3^{+0.3}_{-0.2}$\\
F03068$-$5346 & C & $147.1^{+205.1}_{-19.3}$ & $2.5^{+0.4}_{-0.5}$ & $-0.9^{+0.1}_{-0.1}$ & $0.1^{+0.0}_{-0.1}$ & - & - & - & - & $0.40^{+0.16}_{-0.08}$ & $49.7^{+3.0}_{-3.4}$ & $1.2^{+0.2}_{-0.1}$\\
F03481$-$4012 & PL & $15.0^{+0.3}_{-0.4}$ & - & $-0.8^{+0.0}_{-0.0}$ & - & - & - & - & - & $0.29^{+0.12}_{-0.09}$ & $47.3^{+3.8}_{-3.0}$ & $1.5^{+0.3}_{-0.3}$\\
F04063$-$3236 & C2 1SA & $49.4^{+4.4}_{-4.0}$ & $0.2^{+0.3}_{-0.1}$ & $-1.3^{+0.1}_{-0.1}$ & $0.3^{+0.0}_{-0.0}$ & $5.2^{+0.4}_{-0.4}$ & $0.2^{+0.2}_{-0.1}$ & - & $6.5^{+0.6}_{-0.7}$ & $0.25^{+0.11}_{-0.08}$ & $49.5^{+4.2}_{-3.7}$ & $1.3^{+0.3}_{-0.2}$\\
F06021$-$4509 & C2 & $25.8^{+8.9}_{-8.8}$ & $0.4^{+0.4}_{-0.3}$ & $-1.1^{+0.2}_{-0.2}$ & $0.4^{+0.3}_{-0.1}$ & $5.0^{+1.2}_{-1.0}$ & $0.4^{+0.3}_{-0.3}$ & $-1.3^{+0.1}_{-0.1}$ & $4.4^{+0.9}_{-0.7}$ & $0.09^{+0.01}_{-0.01}$ & $59.8^{+1.6}_{-1.8}$ & $1.1^{+0.2}_{-0.1}$\\
F06035$-$7102 & C2 1SAN & $23.2^{+3.6}_{-2.7}$ & $0.3^{+0.4}_{-0.2}$ & $-1.2^{+0.0}_{-0.0}$ & - & $349.8^{+42.2}_{-38.6}$ & $0.3^{+0.4}_{-0.3}$ & - & $0.4^{+0.0}_{-0.0}$ & $0.65^{+0.12}_{-0.11}$ & $49.3^{+2.1}_{-1.7}$ & $1.1^{+0.1}_{-0.0}$\\
F06206$-$6315 & C2 1SA & $49.0^{+7.4}_{-8.4}$ & $0.6^{+0.6}_{-0.4}$ & $-1.3^{+0.2}_{-0.1}$ & $0.5^{+0.1}_{-0.1}$ & $12.8^{+0.9}_{-1.1}$ & $0.5^{+0.4}_{-0.4}$ & - & $4.5^{+0.4}_{-0.4}$ & $0.67^{+0.13}_{-0.10}$ & $45.9^{+1.8}_{-1.8}$ & $1.1^{+0.1}_{-0.1}$\\
F18582$-$5558 & C2 & $213.2^{+191.4}_{-85.6}$ & $0.3^{+0.4}_{-0.2}$ & $-0.9^{+0.1}_{-0.1}$ & $0.0^{+0.0}_{-0.0}$ & $5.9^{+1.0}_{-0.9}$ & $0.1^{+0.1}_{-0.1}$ & $-1.3^{+0.2}_{-0.1}$ & $5.4^{+0.7}_{-0.9}$ & $0.48^{+0.07}_{-0.08}$ & $43.4^{+1.5}_{-1.4}$ & $1.8^{+0.1}_{-0.2}$\\
F20117$-$3249 & C2 1SAN & $7.3^{+1.7}_{-1.3}$ & $0.6^{+0.6}_{-0.4}$ & $-1.1^{+0.1}_{-0.1}$ & - & $73.0^{+3.4}_{-2.9}$ & $0.5^{+0.6}_{-0.4}$ & - & $1.6^{+0.2}_{-0.2}$ & $1.14^{+0.57}_{-0.38}$ & $37.1^{+2.5}_{-2.1}$ & $1.7^{+0.2}_{-0.3}$\\
F20445$-$6218 & C & $62.0^{+106.3}_{-15.7}$ & $0.6^{+0.5}_{-0.4}$ & $-0.8^{+0.1}_{-0.1}$ & $0.2^{+0.1}_{-0.2}$ & - & - & - & - & $0.40^{+0.24}_{-0.10}$ & $46.6^{+3.1}_{-3.9}$ & $1.3^{+0.3}_{-0.2}$\\
F21178$-$6349 & C & $28.6^{+9.1}_{-7.0}$ & $0.9^{+0.2}_{-0.2}$ & $-1.2^{+0.1}_{-0.1}$ & $0.4^{+0.1}_{-0.1}$ & - & - & - & - & $0.23^{+0.15}_{-0.08}$ & $48.2^{+4.6}_{-4.4}$ & $1.5^{+0.4}_{-0.3}$\\
F21292$-$4953 & PL & $21.0^{+0.6}_{-0.7}$ & - & $-0.5^{+0.0}_{-0.0}$ & - & - & - & - & - & $0.46^{+0.25}_{-0.13}$ & $46.7^{+3.5}_{-3.6}$ & $1.4^{+0.4}_{-0.3}$\\
F21295$-$4634 & C2 1SAN & $1.1^{+0.7}_{-0.6}$ & $0.3^{+0.2}_{-0.2}$ & $-1.0^{+0.1}_{-0.1}$ & - & $31.3^{+7.7}_{-6.1}$ & $0.3^{+0.3}_{-0.2}$ & - & $0.5^{+0.2}_{-0.1}$ & $1.27^{+0.54}_{-0.44}$ & $39.0^{+2.9}_{-2.2}$ & $1.6^{+0.2}_{-0.3}$\\
F23389$-$6139 & C2 1SA & $421.7^{+14.4}_{-14.6}$ & $1.0^{+0.6}_{-0.7}$ & $-1.4^{+0.0}_{-0.0}$ & $0.4^{+0.0}_{-0.0}$ & $91.9^{+6.8}_{-6.5}$ & $0.7^{+0.6}_{-0.4}$ & - & $3.0^{+0.2}_{-0.2}$ & $0.86^{+0.26}_{-0.23}$ & $44.7^{+2.7}_{-2.3}$ & $1.3^{+0.3}_{-0.2}$\\
\bottomrule
\end{tabular}
\end{table*}

Nominal model parameters and their one sigma uncertainties, taken from the posterior distribution  constructed by \textsc{emcee}, are shown in Table~\ref{Table:param}. Using the sampled posterior distribution, we take the 50$^{\mathrm{th}}$ percentile as the nominal value, and the 16$^{\mathrm{th}}$ and 84$^{\mathrm{th}}$ represent the one sigma uncertainties. These posterior distributions were also saved and used when estimating derived quantities, including luminosities or thermal fractions, to accurately propagate errors and maintain covariance amongst a given models fitted parameters. 

An example of a final SED is presented in Fig.~\ref{fig:sed4}, with the remainder presented in the Appendix. These SEDs include the most preferred model judged strictly by the Bayes evidence values. We include model specific features where possible as additional overlaid components. Highlighted regions of all plotted components represent the one sigma (68\%) confidence interval. 


\begin{figure}
\includegraphics[width=\linewidth]{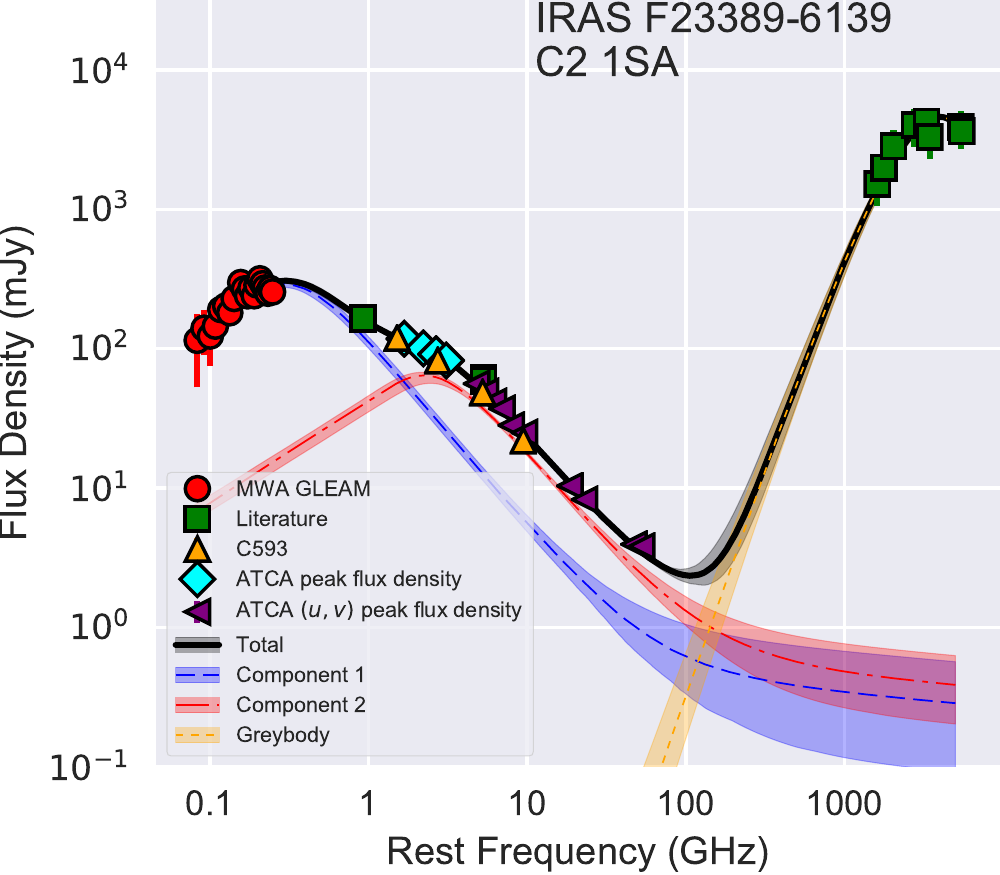}
\caption{The observed data and preferred spectral energy distribution modelling of \iras{F23389$-$6139}. The overlaid model exhibits two distinct FFA turnovers. Highlighted regions represent the 1$\sigma$ uncertainty sampled by \emcee. \label{fig:sed4}}
\end{figure}


\begin{figure}
	\includegraphics{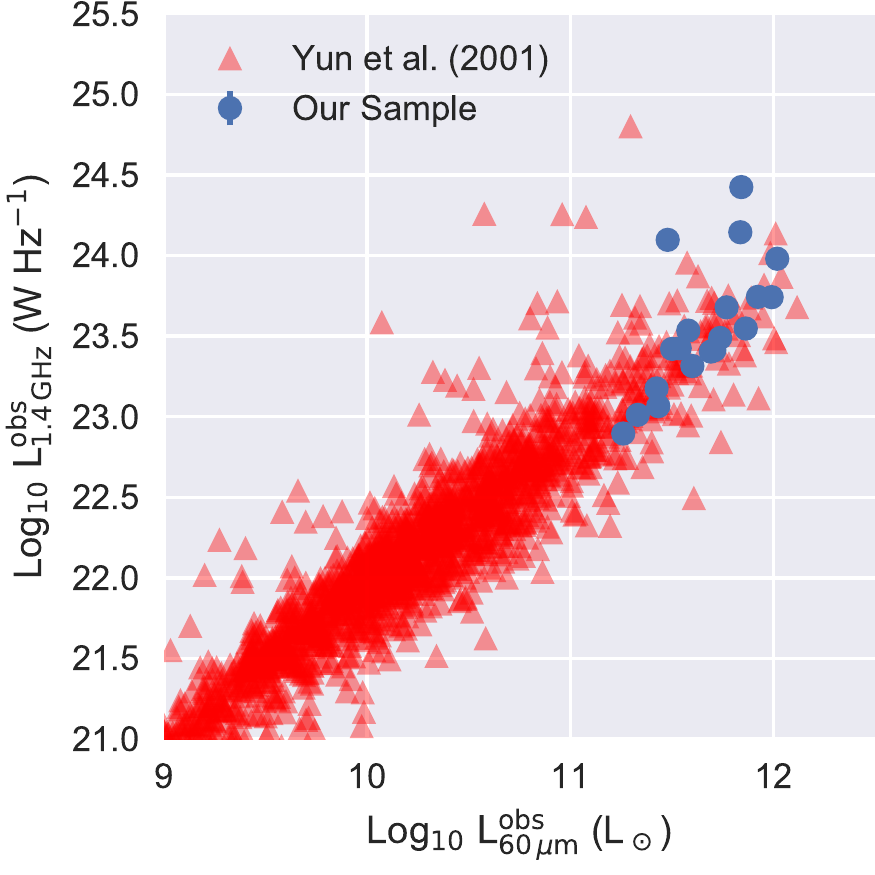}
	\caption{A comparison of the 60\,$\mu$m and 1.4\,GHz luminosities of our 19 source sample and the \citet{2001ApJ...554..803Y} sample. Luminosities have been estimated in the observed frame with no $k$-correction applied. \label{fig:firrc}}
\end{figure}

To examine how our sample resided on the FRC we compared it to the sample of 1809 objects of \citet{2001ApJ...554..803Y}. The \citet{2001ApJ...554..803Y} sample cross referenced the {\em IRAS} 2\,Jy sample with the NRAO VLA All Sky Survey \citep[NRAO;][]{1998AJ....115.1693C} to investigate the FRC over many orders of magnitude. To remain consistent with their work no $k$-correction was applied. We see in Fig.~\ref{fig:firrc} that our sample is consistent with the trend seen by \citet{2001ApJ...554..803Y}, where only three of our objects (\iras{F06035$-$7102}, \iras{F20117$-$3249} and \iras{F23389$-$6139}) have a slightly elevated 1.4\,GHz luminosity. The $q$ parameter, which is the logarithmic ratio between the far infrared flux and 1.4\,GHz flux density of an object, is a further useful illustration of the FRC, where $q$ is defined as:

\begin{dmath}
\label{eq:q}
\\
q \equiv \mbox{log}_{10}\left( \frac{\mbox{FIR}}{3.75 \times 10^{12} \mbox{ W m}^{-2}}\right) - \\ \mbox{log}_{10}\left( \frac{S_{1.4\  \mathrm{GHz}}}{\mbox{W m}^{-2} \mbox{ Hz}^{-1}}\right).
\\
\end{dmath}

\noindent $S_{1.4\,\mathrm{GHz}}$ is the flux density at $\nu=1.4$\,GHz, and FIR is defined as

\begin{figure}
	\includegraphics[width=\linewidth]{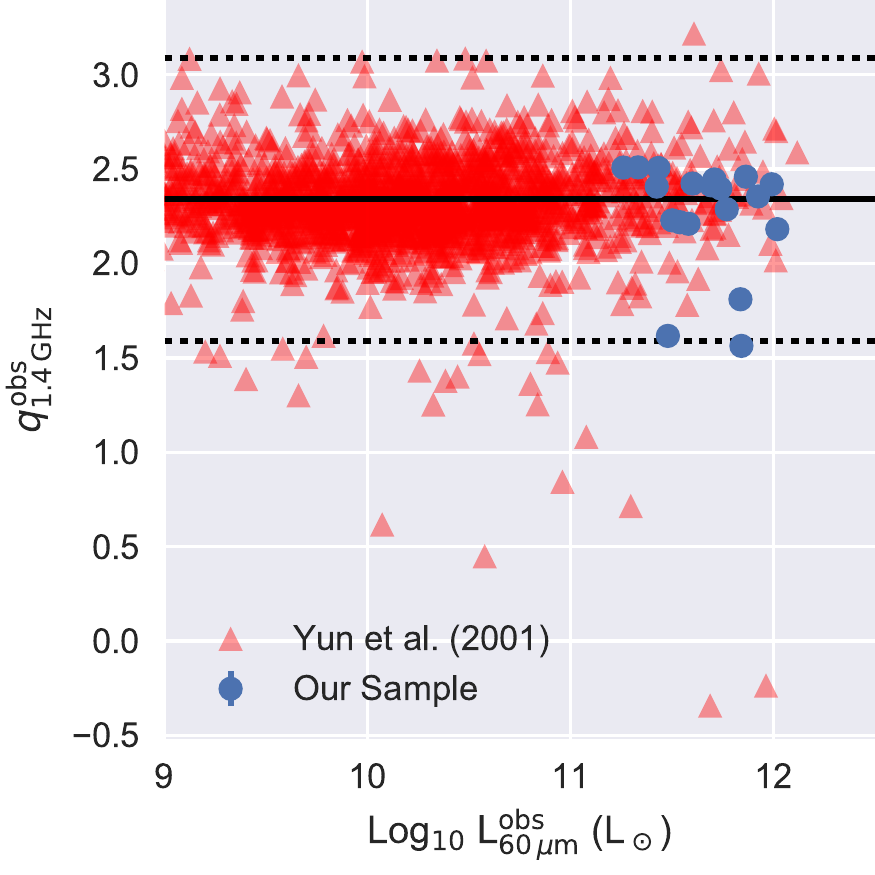}
	\caption{The FRC, as parameterised by the $q$ parameter, of our sample and \citet{2001ApJ...554..803Y}.  The solid horizontal line represents the mean $q = 2.34$, as calculated by \citet{2001ApJ...554..803Y}. The dotted lines represent the radio-excess ({\em below}) and IR-excess ({\em above}) objects, which we defined as three times the standard deviation (SD) of $q$ (SD~$=~0.25$) from the \citet{2001ApJ...554..803Y} sample. \label{fig:firrc-q}}
\end{figure}

\begin{dmath}
\label{eq:fir}
\\
\mathrm{FIR} \equiv 1.26 \times 10^{-14}\left(2.58S_{60\ \mu\mathrm{m}} + S_{100\ \mu\mathrm{m}}\right) \mathrm{  W\ m}^{-2},
\\
\end{dmath}

\noindent where $S_{60\,\mu\mathrm{m}}$ and $S_{100\,\mu\mathrm{m}}$ are the 60 and 100\,$\mu$m band flux densities from {\em IRAS} in Jy \citep{1985ApJ...298L...7H}. The mean $q$ value between 60\,$\mu$m and 1.4\,GHz is typically taken as 2.34 for star formation galaxies \citep{2001ApJ...554..803Y}. Any deviation from this value can be a critical diagnostic of the physical processes driving some object. IR-excess sources ($q>3$) may be highly obscured compact starburst galaxies or dust-enshrouded active galactic nuclei (AGN). Radio-excess objects ($q<1.6$) are caused by excess radio emission originating from an AGN component in the galaxy \citep{2001ApJ...554..803Y}. Some of the dispersion may be influenced in part by variation in extinction and dust temperature, as well as varying timescales associated with different SFR indicators. We show in Fig.~\ref{fig:firrc-q} the distribution of the $q$ parameters as a function of 60\,$\mu$m luminosity and highlight the regions which radio or infrared excess sources occupy.

Of our sample objects IRAS\,F20117$-$3249 and IRAS\,F23389$-$6139 have $q$ values (1.61 and 1.54 respectively) that are approaching the radio-excess region, indicating the potential presence of AGN activity in the observed 1.4\,GHz radio continuum (see Fig.~\ref{fig:firrc-q}). There is no classification of IRAS\,F20117$-$3249 available in the literature, although it has been designated as a galaxy by \citet{2003A&A...412...45P}. IRAS\,F23389$-$6139, however, has been classified as a starburst based on optical imagery \citep{1997A&AS..124..533D} and infrared template modelling \citep{2003MNRAS.343..585F}. \iras{F06035-7102} also has a slightly elevated $q$ parameter of 1.8. It has been classified in the literature as a starburst based on optical spectral features and infrared modelling \citep{2000MNRAS.317...55S,2003MNRAS.343..585F}.  

\subsection{Thermal Fraction}

The thermal fraction of a source is a measure of how much of the observed radio continuum is comprised of thermal free-free emission. At increasing frequencies, due to the steep spectral index of non-thermal synchrotron emission, the thermal free-free emission begins to dominate the total observed continuum.  H\,\textsc{ii} regions, which are traced by thermal free-free emission, are an excellent probe of current star formation. In the GHz regime free-free emission represents roughly 5 to 10\% of the total radio continuum \citep{1992ARA&A..30..575C,2013ApJ...777...58M} and due to its flat spectral index ($S_\nu \propto \nu^{-0.1}$) it is relatively difficult to isolate. The broad coverage of our radio continuum SEDs however allows us to investigate this property. For each source, using the best fit model, we compute the total amount of thermal emission in order to derive appropriate nominal thermal fraction values. 
 
\begin{figure}
\includegraphics{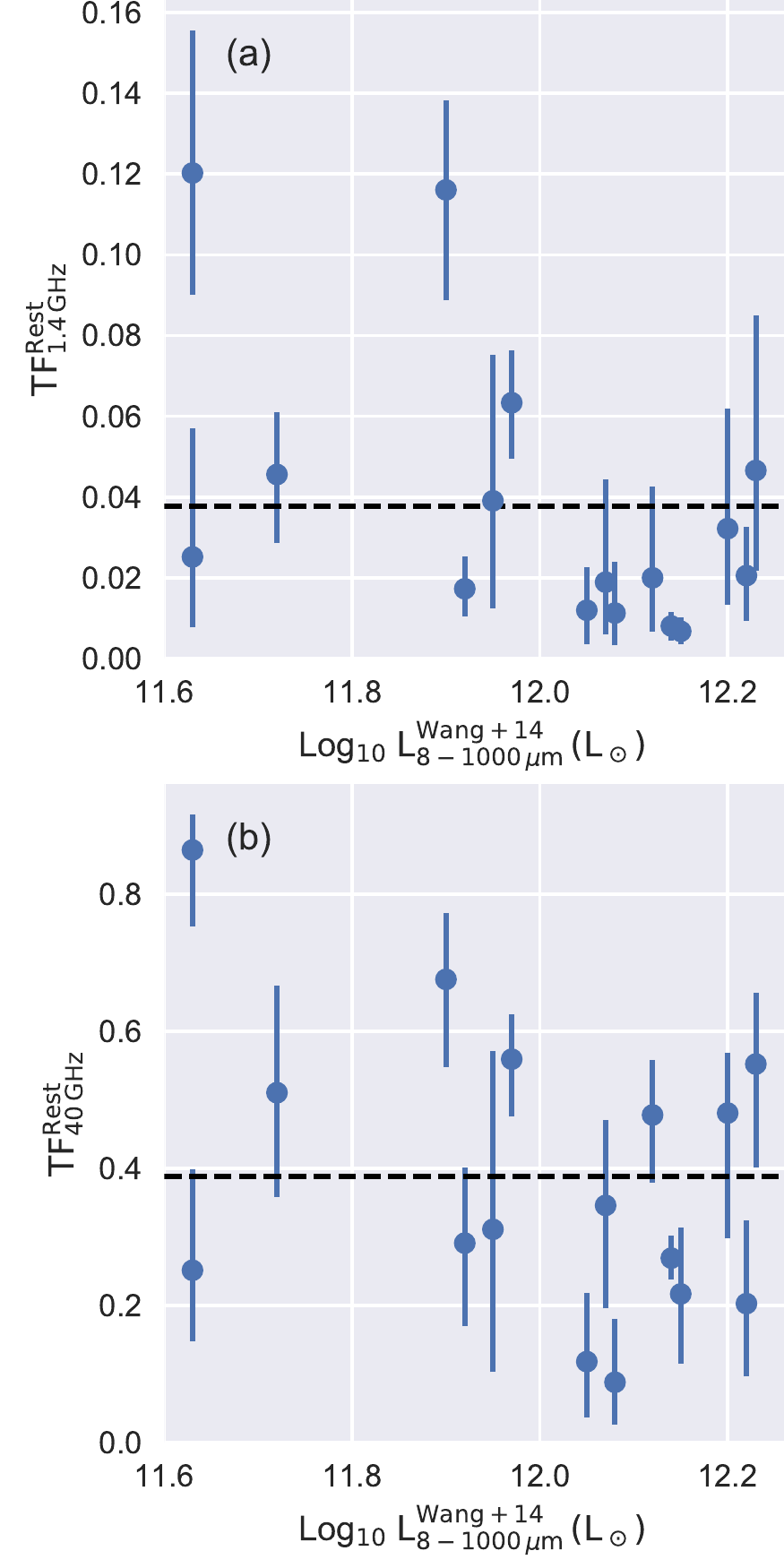}
\caption{A comparison between the total infrared emission and the estimated thermal fraction of the rest frame at 1.4\,GHz \textit{(a)} and at 40\,GHz \textit{(b)}. Dashed horizontal lines represent the average thermal fraction. \label{fig:tir-tf}}	
\end{figure}

We find at low frequencies the thermal fraction makes up only a small fraction of the total radio continuum emission. This is similar to earlier studies \citep{1990ApJ...357...97C,1992ARA&A..30..575C,1992ApJ...401...81P}, where at 1.4\,GHz the typical thermal fraction was estimated to be around 10\%. In Fig.~\ref{fig:tir-tf}a we show that at 1.4\,GHz the estimated thermal fraction is fairly constant at around 3 to 10\%, with the average thermal fraction being 3.8\%. This is in line with \citet{2013ApJ...777...58M}, who find in a sample of 31 local starburst galaxies that the thermal fraction at 1.4\,GHz is $\approx5$\%. At 40\,GHz (Fig.~\ref{fig:tir-tf}b), the thermal fraction makes a much larger contribution to the modelled radio continuum, ranging between 35 to 80\% with an average of 38.8\%. 

These thermal fractions, in principal, could be effected by missing interferometric flux at higher frequencies where free-free emission processes begin to become more dominate. However, we do not expect this to be an issue as our brightness temperature estimates (see \S\ref{sec:bt}) are above the lower limit for a face-on spiral galaxy and only approach this limit at the highest frequency in Q-band. 

\subsection{Spectral curvature and emission measures}

Similar to \citet{2010MNRAS.405..887C}, the radio continuum SEDs in our sample of objects are rarely characterised well by a simple power law. The broad frequency range covered by our data shows the presence of multiple bends or turnovers, which we attribute to the effects of FFA. Low-frequency data from the MWA GLEAM data shows clear cases of low frequency turnovers, as illustrated well by \iras{F01388$-$4618} and \iras{F23389$-$6139}. At higher frequencies we see in a subset of our sources evidence supporting a `kink' in the radio continuum spectrum. Likewise to the turnover at low frequency, we attribute this to a secondary FFA component with a higher optical depth.   

Four objects from our sample had a evidence value that most supported a `simple' model (a power law or the simple normalisation of synchrotron and free-free power law components). Of these four, objects only \iras{F01419$-$6826} had a competing higher order  model. The remaining 15 objects all had higher order (i.e. turnover due to FFA) models most supported by the evidence, where only source \iras{F03068$-$5346} had a `simple' competing model. 

A common feature seen in our SEDs is the steepening of the radio continuum spectrum between the 4 to 10\,GHz regime. A similar effect was also seen by \citet{2008A&A...477...95C,2010MNRAS.405..887C} and \citet{2011ApJ...739L..25L}.  In cases where the MWA GLEAM low frequency measurements indicates a low frequency turnover, this steepening is often modelled by an additional component of FFA attenuated synchrotron and free-free emission. This higher order complexity is supported by both an improved $\chi^2$ statistic and $\Zagr$ value.

The turnover frequency due to FFA is dependent on where the optical depth reaches unity. Generally it is assumed that the emitting \htwo\ regions form a cylinder orientated along of line of sight with constant temperature and density \citep{1992ARA&A..30..575C}. In such scenarios, the free-free opacity is well approximated by:

\begin{equation}
\tau_\nu = 3.28\times10^{-7}\left(\frac{T_e}{10^4\mathrm{\,K}}\right)\left(\frac{\nu}{\mathrm{GHz}}\right)^{-2.1}\left(\frac{EM}{\mathrm{pc\,cm}^{-6}}\right),
\end{equation}
 
\noindent where $T_e$ is the electron temperature of the \htwo\ emitting region, typically taken as $10^4$\,K, and EM is the emission measure, defined as:

\begin{equation}
\label{eq:em}
	\frac{EM}{\mathrm{pc\, cm}^{-6}} = \int_{\mathrm{los}} \left(\frac{N_e}{\mathrm{cm}^{-3}}\right)^2 d\left(\frac{s}{\mathrm{pc}}\right).
\end{equation}

EM is the integral of the electron density along the line of sight of a \htwo\ region of depth $s$. Using the above form, for frequencies above the  turnover frequency $\nu_t$, the free-free spectrum follows a power law of $\alpha\sim-0.1$. Once the optical depth reaches unity, the spectrum transitions to  the Rayleigh-Jeans law, described well by $\nu^2$. Using the turnovers constrained by our modelling, we have estimated the EMs of our sources, outlined in Table~\ref{Table:em}, using Equation~\ref{eq:em}. We label the corresponding EM of $\nu_{t,1}$ and $\nu_{t,2}$ for all models as EM$_1$ and EM$_2$ respectively.

\tabcolsep=0.2cm
\begin{table}
\centering
\caption{An overview of the emission measures (EM) derived for each source from the model most supported by the evidence. Objects without an emission measure constrained are marked by `--'. \label{Table:em}}
\begin{tabular}{cll}
\toprule
Source & EM$_1$ & EM$_2$  \\
 {\em IRAS} & $10^6$\,cm$^{-6}$\,pc & $10^6$\,cm$^{-6}$\,pc \\
\midrule
F00198-7926 & 0.016 & 13.851\\
F00199-7426 & -- & 0.067\\
F01388-4618 & 0.021 & --\\
F02364-4751 & 0.017 & --\\
F03068-5346 & 0.003 & --\\
F04063-3236 & 0.029 & 15.762\\
F06021-4509 & 0.038 & 6.88\\
F06035-7102 & -- & 0.044\\
F06206-6315 & 0.059 & 7.1\\
F18582-5558 & -- & 10.599\\
F20117-3249 & -- & 0.858\\
F20445-6218 & 0.01 & --\\
F21178-6349 & 0.043 & --\\
F21295-4634 & -- & 0.08\\
F23389-6139 & 0.044 & 3.11\\
\bottomrule
\end{tabular}
\end{table}

For systems with multiple intense starburst regions that have been integrated over by a large synthesised radio-telescope beam, their superposition of radio continuum features will form the observed SED. The orientation of such regions will play a crucial role in the spectral curvature across a broad frequency range. Regions which are small and deep will posses much higher EMs than those which are more widespread and shallow relative to our observing angle. Although the EM is tied to the spatial size of an object, which can vary as a function of frequency with increasing amounts of diffuse synchrotron, we have no evidence to suggest we are resolving our sample, particularly at high frequency where we have obtained critical short spacing data. 

\subsection{Far-infrared to radio correlation}

The radio continuum emission is considered an ideal tracer of star formation as it is not effected by dust attenuation. In terms of the local Universe ($z < 0.2$) it is fairly well calibrated by bootstrapping the radio continuum SFR against the far-infrared SFR via the FRC. 

Although understood well in the local Universe, it is unknown whether the FRC will evolve with increasing redshift. As outlined by \citet{2009ApJ...706..482M}, due to the changing composition of the radio continuum with increasing frequency (which is what would be doppler shifted to lower frequency) and synchrotron suppression effects that scale with $(1+z)^4$ caused by inverse-Compton losses, it is thought that the FRC will need to be `recalibrated' to be compatible with the high redshift Universe. \cite{2011ApJ...731...79M} however see no evidence of evolution in the FRC up to $z\sim2$ using image stacking techniques, suggesting that the FRC is more physically complex than first thought. As we show in Fig.~\ref{fig:firrc}, our sample of objects follow the FRC. 

\begin{figure*}
\includegraphics[width=\linewidth]{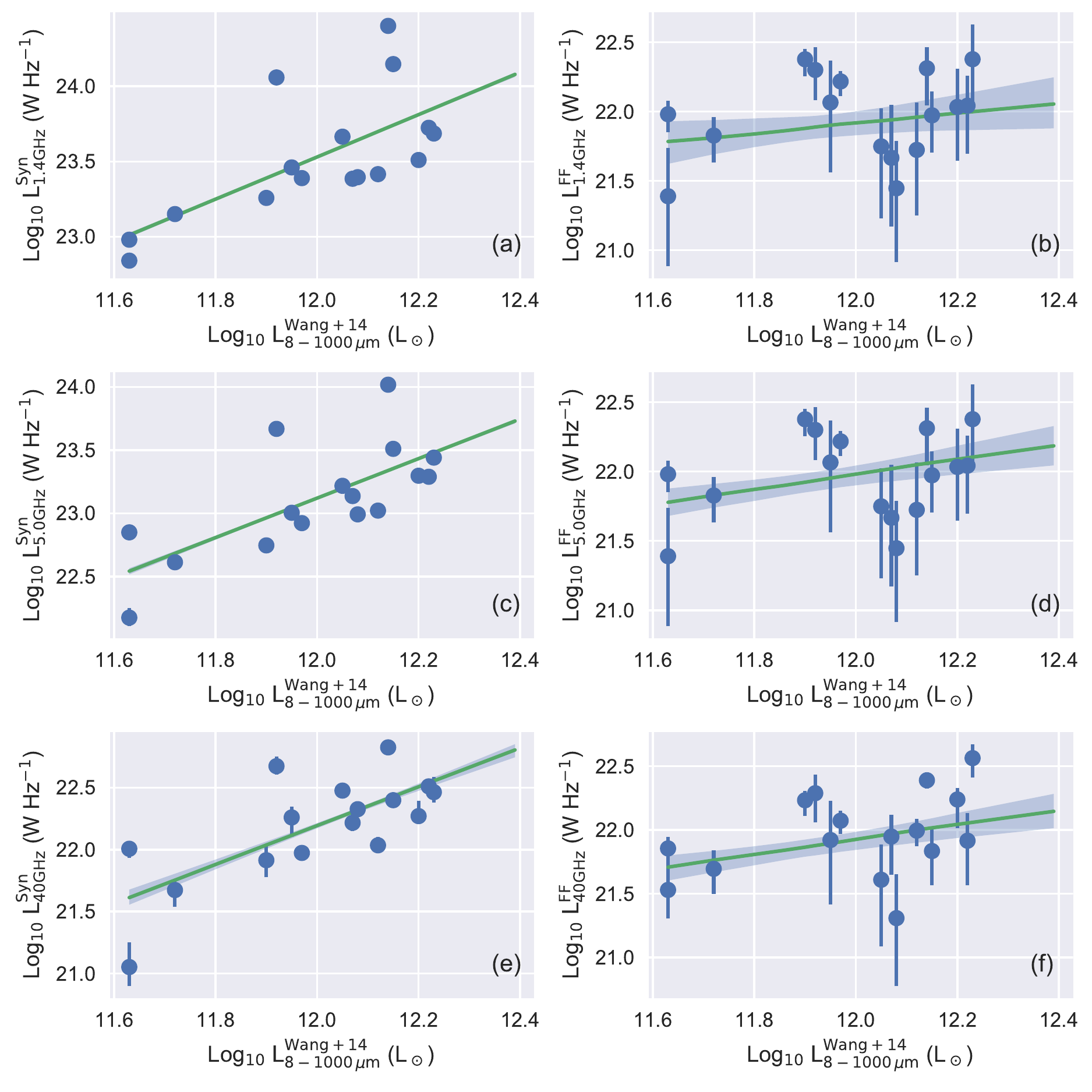}
\caption{A comparison between the total infrared derived luminosity, as presented by \citet{2014MNRAS.442.2739W}, and the constrained synchrotron \textit{(a,c,e)} and free-free luminosities \textit{(b,d, f)} at 1.4\,GHz, 5.0\,GHz and 40\,GHz of sources in our sample. The green line and its highlighted region represents a non-weighted linear regression and the corresponding 1$\sigma$ uncertainty region determined from 1000 realisations.  \label{fig:fir_v_SynandFF}}	
\end{figure*}

The synchrotron and free-free emission mechanisms that make up the radio continuum are both tracers of star formation across different timescales. Given the posterior distribution sampled by \emcee\ of the most supported model of each source, we compare in Fig.~\ref{fig:fir_v_SynandFF} the total far-infrared \citep[taken from ][]{2014MNRAS.442.2739W} against the decomposed synchrotron and free-free components at 1.4 and 40\,GHz. For each comparison we also include the results of a non-weighted linear fit against multiple realisations (N=1000) of our data, drawn randomly from the posterior distribution. Highlighted regions represent the 1$\sigma$ uncertainty of the best fit parameters of this process. 

The total far-infrared correlates well with the modelled synchrotron luminosity for all sources at 1.4\,GHz, as demonstrated in Fig.~\ref{fig:fir_v_SynandFF}a. This can simply be attributed to synchrotron emission dominating the radio continuum at 1.4\,GHz \citep{1992ARA&A..30..575C,2001ApJ...554..803Y,2003ApJ...586..794B,2006ApJ...651L.111M}. The two outlying objects, whose synchrotron luminosities are in excess of $10^{24}$\,W\,Hz$^{-1}$ are IRAS\,F20117$-$3249 and IRAS\,F23389$-$6139. 

Free-free emission is a more reliable probe of SFR with these increasing redshifts as it directly tracers H\textsc{ii} regions ionised by nearby high mass stars and is unaffected by inverse-Compton losses. Identifying the free-free emission at low frequencies, where it contributes $\sim5-10$\% at 1.4\,GHz, is difficult and few studies have successfully isolated its signature \citep{1992ApJ...401...81P,2010MNRAS.405..887C,2010ApJ...709L.108M,2012ApJ...761...97M,2016MNRAS.461..825G}. This is demonstrated in the top panel of Fig.~\ref{fig:fir_v_SynandFF}b where there is considerable uncertainty associated with the constrained free-free luminosity at 1.4\,GHz. 
  
With increasing frequencies, there is a change in the composition of the radio continuum. Synchrotron emission, due to its steep spectral index, quickly begins to weaken. We show in Fig.~\ref{fig:fir_v_SynandFF}c-e that though there is still a strong correlation between the total infrared and the estimated synchrotron luminosity, it is one with increased uncertainty when compared to the equivalent relation constrained at 1.4\,GHz (Fig.~\ref{fig:fir_v_SynandFF}a). The correlation between the total infrared and free-free luminosity at 40\,GHz (Fig.~\ref{fig:fir_v_SynandFF}f) is far more constrained than it was at 1.4 and 5.0\,GHz (Fig.~\ref{fig:fir_v_SynandFF}b-d).
 
In Table~\ref{Table:frc} we list the best fit values from a simple linear regression between the total infrared (in units of L$_\odot$) and the decomposed synchrotron and free-free luminosity components at 1.4, 5.0 and 40\,GHz (in units of W\,Hz$^{-1}$). Errors were estimated by drawing 1000 realisations of the luminosities from the posterior distribution sampled by \emcee. We find these results acceptable given that we have less than one order of magnitude of range in the total infrared luminosities. 

{\tabcolsep=0.2cm
\begin{table}
\centering
\caption{The fitted gradient and normalisation components of a non-weighted linear fit between the total infrared and decomposed radio continuum luminosities. Synchrotron and free-free luminosity components are labeled as `Syn' and `FF' respectively. We also provide the mean and standard deviation of the $q$ parameter, derived using the total infrared luminosity from \citet{2014MNRAS.442.2739W}, for each of the correlations.  \label{Table:frc}}
\begin{tabular}{cccccc}
\toprule
$\nu$ & Emission & Gradient & Norm. & $q\pm\sigma_q$ \\
 (GHz) &&& (Log$_{10}\mathrm{\,W\,Hz}^{-1}$) &  \\
\midrule
\rule{0pt}{3ex} 1.4 & Syn & $1.40^{+0.03}_{-0.04}$ & $6.68^{+0.47}_{-0.41}$ & 2.53$\pm$0.38\\
\rule{0pt}{3ex} 5.0 & Syn & $1.56^{+0.06}_{-0.05}$ & $4.36^{+0.62}_{-0.66}$ & 2.94$\pm$0.40\\
\rule{0pt}{3ex} 40.0 & Syn & $1.57^{+0.13}_{-0.15}$ & $3.39^{+1.77}_{-1.56}$ & 3.87$\pm$0.39\\
\rule{0pt}{3ex} 1.4 & FF & $0.33^{+0.43}_{-0.33}$ & $17.95^{+3.92}_{-5.10}$ & 4.09$\pm$0.33\\
\rule{0pt}{3ex} 5.0 & FF & $0.53^{+0.27}_{-0.25}$ & $15.60^{+2.93}_{-3.27}$ & 4.04$\pm$0.34\\
\rule{0pt}{3ex} 40.0 & FF & $0.58^{+0.25}_{-0.23}$ & $14.97^{+2.74}_{-2.97}$ & 4.09$\pm$0.34\\
\bottomrule
\end{tabular}
\end{table}
}

\citet{1992ApJ...401...81P} performed a similar analysis for a sample of 31 galaxies. Their study used a single model equivalent to Eq.~\ref{eq:synandff} and found that the decomposed synchrotron and free-free components are tightly correlated to the far-infrared across roughly three orders of magnitude. At 5.0\,GHz, they estimate the gradient of the synchrotron-FIR and free-free-FIR correlations to be $1.33\pm0.1$ and $0.93\pm0.02$ respectively. These are comparable to the correlations derived above, particularly the synchrotron-TIR component at $\nu=5.0$\,GHz. Although we are using the total infrared luminosities, defined as the bolometric luminosity from 8 to 1000\,$\mu$m, derived by \citet{2014MNRAS.442.2739W} and their IR template fitting routines, the bulk of emission for star forming galaxies in this regime is emitted in the FIR \citep{1988ApJS...68..151H,1992ARA&A..30..575C}. This difference would influence the normalisation component which are not being compared here. 

We acknowledge that these correlations may be partly a result of our sample selection criteria. By ensuring that sources were selected such that their was no radio or infrared excess objects, as measured by deviation of their $q$ parameter, there may be a selection bias. If sources were purposely selected to be on the FRC, then the components of the radio continuum modelling are also likely to follow similar trends. However, the initial constraints on $q$ where broad enough to be considerably larger then the intrinsic scatter in the original correlation (see Fig.~\ref{fig:firrc-q}) and those reported here.

\section{Discussion}
\subsection{Spectral Curvature - Physical Origin?}
	Sixteen objects in our sample show spectral characteristics that are not consistent with a simple power law model. An inconsistent flux calibration scale may also influence spectral features when comparing data across a broad frequency range. For ATCA data, PKS\,1934$-$638 is almost exclusively used as a flux calibrator. This gigahertz peaked spectrum source has been well characterised from low to high frequency, and is estimated to be absolute spectrum of 3C286 and 3C295 on the \citet{1977A&A....61...99B} scale \citep{1934,0004-637X-821-1-61}. As it was used to provide a flux calibration scale for all ATCA data from 2.1 to 25\,GHz for sources in our sample, it is unlikely that high frequency kinks between 4 to 10\,GHz are due to mis-matched matched flux scale. 

At the time of observing the high frequency Q band data, Uranus was the preferred flux calibrating source at ATCA. The flux density accuracy at these frequencies is estimated to be within 10\%. An over-estimated flux density at these frequencies would have the effect of increasing the amount of free-free emission while model fitting, producing a more pronounced flattening at higher frequencies. Therefore, we have added an additional 10\% error in quadrature as a measure to counteract this effect. Low frequency data from SUMSS and MWA GLEAM DR1 both use the Molonglo reference catalog \citep[MRC; ][]{1981MNRAS.194..693L,1991Obs...111...72L} in large part to craft a flux calibration scale that is accurate to 2-3\% on the \citet{1977A&A....61...99B} scale. To account for any residual flux calibration mis-match, we inject an additional 5\% error in quadrature for all flux density measurements obtained through NED, including measurements from SUMSS, or archived ATCA observations (Table~\ref{atoa-data}). 

Therefore, the curvature features we see in our modelling, we believe, are physical in origin. When the derived emission measures (Table~\ref{Table:em}) are compared to similar studies we find that they are consistent. \citet{2010MNRAS.405..887C} studied a sample of 20 luminous and ultra luminous infrared galaxies using data from 244\,MHz to frequencies in excess of 23\,GHz. Although their SEDs are more sparsely sampled than those in this study, they find evidence that suggests multiple FFA components with varying optical depths. They find emission measures in the range of 0.12 to 140\,$\times10^6\,\mathrm{cm}^{-6}\,\mathrm{pc}$. Similarly, observations using MERLIN of compact sources in M82 at 408\,MHz presented by \cite{1997MNRAS.291..517W} are in agreement of emission measures derived from low frequency turnovers for our sample. At higher frequencies, \citet{2000AJ....120..670N} also find emission measures in excess of $10^8$\,cm$^{-6}$\,pc for compact \htwo\ regions and supernova remnants observed in the NGC\,4038 and NGC\,4039 merger system. 

\begin{figure}
	\includegraphics{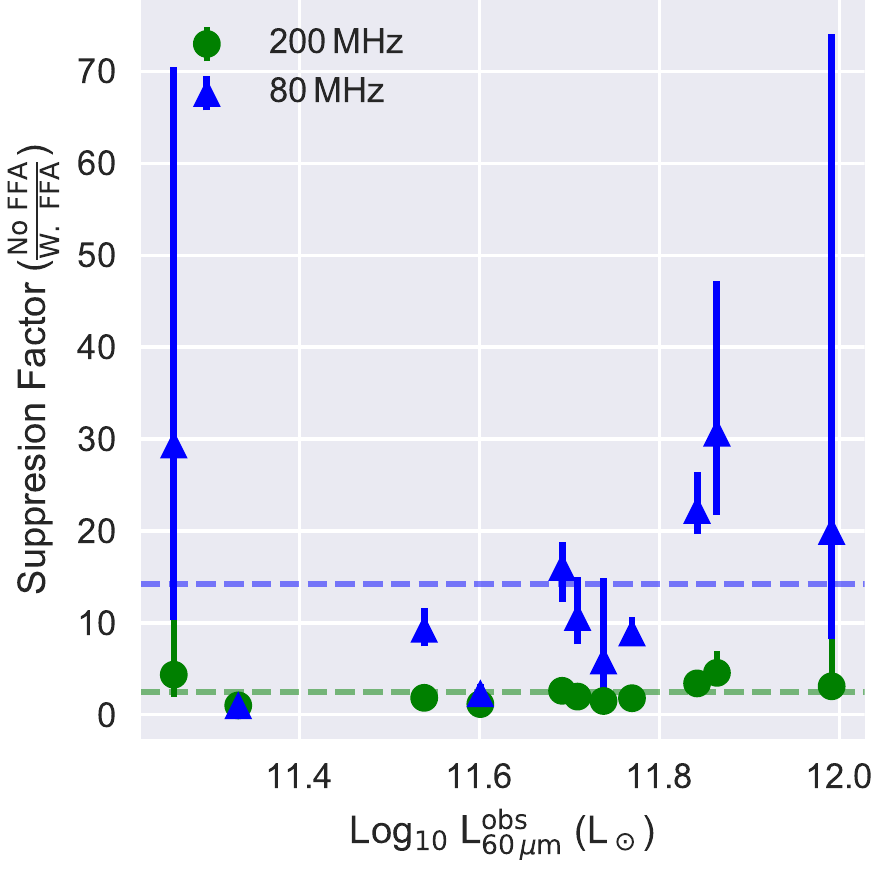}
	\caption{A comparison between the observed 60\,$\mu$m luminosities and the ratio of the observed radio continuum with and without the effects of FFA at 80 and 200\,MHz. The dashed horizontal lines show the average suppression factor for each group. \label{fig:noffa}}
\end{figure}

\arp, the closest ultra luminous infrared galaxy, has observational evidence of a double starburst nucleus thought to be powered by a recent merger \citep{1998ApJ...493L..17S,1998ApJ...507..615D,2011ApJ...729...58E}. Using radio recombination lines to constrain turnover features, \citet{2000ApJ...537..613A} argue that the radio continuum SED of \arp\ is best characterised by three regions of star formation with turn over frequencies at roughly 0.5, 1.4 and 40\,GHz. Using their resolved data allowed them to place further constraints on the emitting size and density of the three emission measures, which they modelled as $1.3\times10^5$, $5.0\times10^6$ and  $6.3\times10^9$\,cm$^{-6}$\,pc. Recently, \citet{2017arXiv170202434K} performed radio continuum modelling of NGC\,253 from 76\,MHz to 11\,GHz. They found that the galaxy was best described as the sum of a discrete central starburst region, modelled as an internally free-free absorbed synchrotron plasma, with an additional synchrotron component that flattens at low frequency.

The effects of FFA on the radio continuum have be investigated by \citet{2009AJ....137..537L} and \citet{2014AJ....147....5R} on resolved sources embedded within the nearby starburst galaxies NGC\,4945 and NGC\,253. Both studies find evidence of FFA and a range of turnover frequencies between 2 to 10\,GHz. \citet{2009AJ....137..537L} also note that the free-free opacity is highest towards the nucleus of NGC\,4945, but varies significantly. This implies a clumpy composition of star forming regions throughout the system.  

More broadly, resolved multi-wavelength studies of intense starburst galaxies, in the same class as those in our sample, also show multiple, distinct clumps. \citet{2001MNRAS.326.1333F} used the Wide Field Planetary Camera 2 (WFPC2) on the \textit{Hubble Space Telescope (HST)} to study 23 Ultra LIRGs (ULIRGS). They find that most observed sources are in some stage of merger with stellar population synthesis modelling suggesting ages less then several Gyr. Colour maps (based on multiple filters) show a number of distinct `knots' that are clearly distinguished from the surrounding environment which are likely regions of intense starburst activity. Similar \textit{HST} I-band imaging by \citet{2000ApJ...529L..77B} also shows that U/LIRGs are often interacting systems in some stage of merger. 

\subsection{Effects of FFA on low frequency extrapolations}

Euclidean normalised radio source counts of extragalactic objects are a useful cosmological tool \citep{2010A&ARv..18....1D}. Previous to low frequency SKA pathfinder projects and their all sky surveys, including MWA and LOFAR, the radio sky at low frequency was extrapolated from slightly higher frequency surveys and assumed power laws. Although useful as an initial estimate this approach ignores low frequency turnovers caused by free-free absorption. Understanding the behaviour of starburst galaxies, and any deviations from the extraploated optically thin spectrum, will be important for interpreting the well known uptick in the Euclidean normalised radio source counts \citep{2003MNRAS.341L...1G,2008MNRAS.386.1695S}. 
 
In our sample we find that most objects have a well characterised turnover component. In Fig.~\ref{fig:noffa} we compare the effects of omitting this feature and the effects that it may have on simple extrapolation. Flux densities without FFA were obtained by removing the frequency dependent $\tau$ parameter from the most preferred model. We show an example of this in Fig.~\ref{fig:no_ffa}. 

\begin{figure}
	\includegraphics{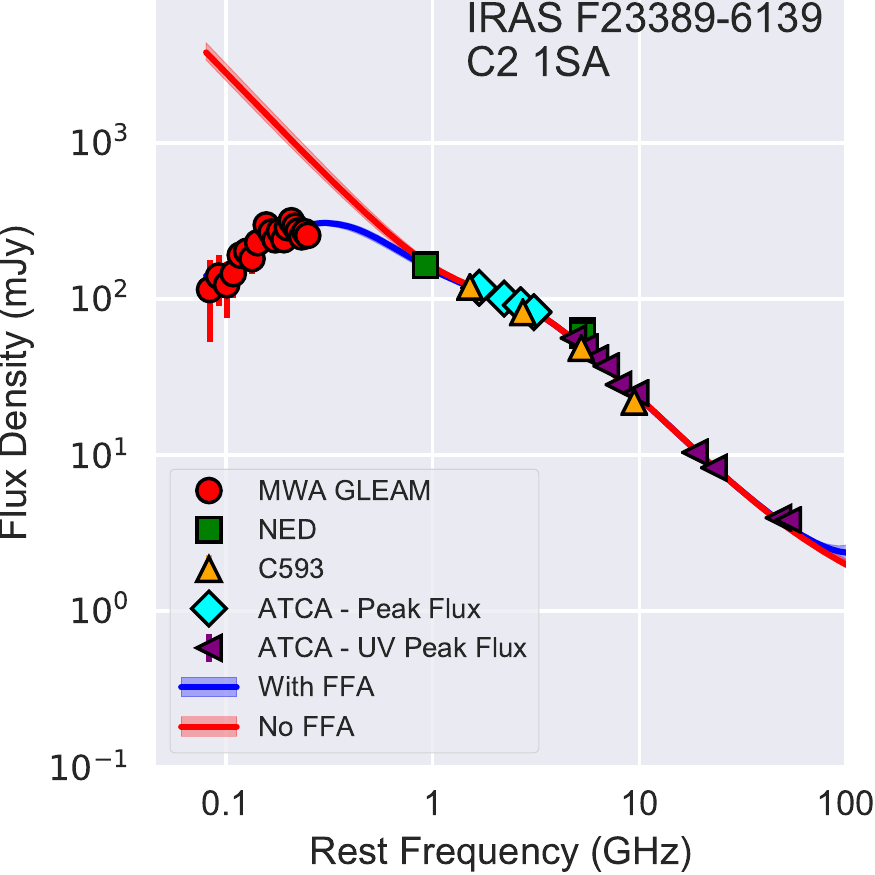}
	\caption{The constrained SED of IRAS\,F23389$-$6139 with the effects of FFA at low frequencies removed. All other constrained parameters are unchanged. \label{fig:no_ffa} }
\end{figure}

At 200\,MHz (Fig.~\ref{fig:noffa}) there is only a small difference to the estimated radio continuum when ignoring the effects of FFA. All but one of our seventeen sources with a low frequency turnover are below a suppression factor of 5, with the average being $\sim2.2$ (ie. without FFA, the radio continuum would be 2.2 times higher then what is observed). At 80\,MHz we find that FFA has a far more significant effect on the observed fluxes. Using the applied modelling we find that at 80\,MHz (Fig.~\ref{fig:noffa}) that the estimated radio continuum without FFA is, on average, $\sim12$ times larger then the observed SED. 

Including curvature due to FFA for SFG type sources when estimating low frequency source counts and confusion limits will be important. Earlier studies have typically used simple power law models with varying spectral indices when extrapolating fluxes between frequencies \citep{2007MNRAS.379.1442W,2016MNRAS.459.3314F}. Similarly, \citet{2008MNRAS.388.1335W} performed a semi-empirical simulation of the radio continuum sky out to $z=20$. In their simulation a turnover due to FFA was assumed at 1\,GHz for all starburst galaxies. Our sample, which is constructed to be representative of high redshift starburst galaxies, have turnovers between 90 to 500\,MHz (if they were detected). Incorporating our turnover frequencies into the assumed model of \citet{2008MNRAS.388.1335W} will have the effect of \textit{increasing} the modelled flux densities at low frequencies. Galvin et al. (in. prep) is exploring the degree of change and how confusion limits could be effected. 

\subsection{Synchrotron Spectral Index}

The mean modelled synchrotron spectral index in our sample is $\alpha=-1.06$, noticeably steeper than the canonical spectral index of $\alpha\sim-0.8$ often assumed for star forming galaxies \citep{1992ARA&A..30..575C}. This is based on the well constrained power-law slope of the distribution of cosmic rays we observe at Earth which directly relates to the synchrotron spectral index. This difference is larger then our modelling uncertainties and seems a real outcome of modelling.  We consider here four mechanisms which could explain the steeper than Milky Way spectral index: (i) electron cooling via inverse-Compton (IC) losses, (ii) ageing of relativistic electrons energy distribution, (iii) a steeper power-law in the relativistic electrons energy distribution or (iv) a different galaxy morphology for extreme starbursts affecting cosmic ray diffusion. We consider (i) to be unlikely as IC losses are weak at low redshift and competing against other losses \citep{2010ApJ...717..196L}. Ageing of the starburst is unlikely as the IR luminosities (which are a more instantaneous tracer of SFR) are still high. 

There is some evidence in the literature from GeV and TeV observations of galaxies with higher star formation rate having a steeper cosmic ray power law index \citep[e.g. ][]{2016arXiv161207290W}. This steeper cosmic ray power law index would then infer a steeper synchrotron spectrum. Our star forming galaxies have even higher SFRs so might be expected to have synchrotron spectral indices as steep as we see here. Interestingly, in the case where a single power law is preferred over multiple components with turn-overs the fitted spectral index is lower and closer to the canonical value. One explanation for this, and the lower spectral indices typically seen in deep surveys \citep{2009MNRAS.397..281I,2012MNRAS.426.2342H}, is that simple 2-3 point broadband spectral indices miss the complexity of spectral structure we observe here and `average' over a steeper power law with multiple turn-over components. 

The final possibility is that powerful stardusts are fundamentally different to lower SFR galaxies previously studied in detail. \citet{2010ApJ...717..196L} describe scenarios where `puffy' starburst galaxies, whose volume density is far less then compact starbursts, can exhibit a steeper cosmic ray distribution index by having a far higher scale height for the cosmic ray distribution. Therefore we suggest that the steep synchrotron spectral indices modelled in our sample of sources are caused by either a steeper cosmic ray distribution index or a physical difference in the nature of these galaxies compared to nearby lower SFR SFGs. 

\section{Conclusion}
We have modelled the radio continuum across a broad frequency range (70\,MHz to 48\,GHz) for a sample of 19 LIRGS, selected specifically to represent the types of objects to be discovered by SKA and its pathfinder projects at high redshifts. We find that:
\begin{enumerate}
	\item in our sample of 19 objects the radio continuum of only three (16\%) sources were characterised well by a single power law component over a broad frequency range;
	\item Eleven objects show evidence of a low frequency turnover ($\nu<800$\,MHz), with ten exhibiting features between 400\,MHz to 6.5\,GHz that are consistent with a higher frequency turnover. Six sources in our sample of 19 show evidence of both a low and mid-to-high frequency turnover. These could be explained by considering free-free absorption processes acting across multiple regions of star formation, each with different optical depths, that are than integrated over by the large radio synthesised beam; 
	\item the intrinsic components that make up the radio continuum are correlated with the far-infrared, with the Syn-IR correlations being steeper then the FF-IR trends, similar to the FRC used to calibrate current 1.4\,GHz SFR tracers; and
	\item without accounting for the effects of FFA the low frequency radio emission of faint starburst galaxies is susceptible to being over estimated by as much as a factor of 30 when using a simple power law scaling from a higher frequency, which may influence estimations of low frequency source counts. 
	\item the mean synchrotron spectral index of our sample is $\alpha=-1.06$, which is steeper then the canonical value of $\alpha=-0.8$. We suggest that this is associated with a steeper cosmic ray distribution index.
\end{enumerate}

In part two of this series of papers we will present the optical integral field spectroscopy obtained for our sample of 19 LIRGS, focusing on the resolved distribution of the Balmer stellar emission lines that trace the same emitting matter as free-free emission, and dust corrected H$\alpha$ star formation rates. 

\section{Acknowledgements}
The authors would like the thank the anonymous referee whose comments and suggestions improved the presentation of this manuscript.
The Australia Telescope Compact Array is part of the Australia Telescope National Facility which is funded by the Commonwealth of Australia for operation as a National Facility managed by CSIRO.  This paper includes archived data obtained through the Australia Telescope Online Archive (\url{http://atoa.atnf.csiro.au}).
This scientific work makes use of the Murchison Radio-astronomy Observatory, operated by CSIRO. We acknowledge the Wajarri Yamatji people as the traditional owners of the Observatory site. Support for the operation of the MWA is provided by the Australian Government (NCRIS), under a contract to Curtin University administered by Astronomy Australia Limited. We acknowledge the Pawsey Supercomputing Centre which is supported by the Western Australian and Australian Governments.
RMcD is a recipient of an ARC Future Fellowship.
Parts of this research were conducted by the Australian Research Council Centre of Excellence for All-sky Astrophysics (CAASTRO), through project number CE110001020
This research has made use of the NASA/IPAC Extragalactic Database (NED) which is operated by the Jet Propulsion Laboratory, California Institute of Technology, under contract with the National Aeronautics and Space Administration.

\bibliographystyle{mn2e}
\bibliography{references}

\begin{thebibliography}{}

\bibitem[\protect\citeauthoryear{{Ambikasaran}, {Foreman-Mackey}, {Greengard},
  {Hogg} \& {O'Neil}}{{Ambikasaran} et~al.}{2014}]{hodlr}
{Ambikasaran} S.,  {Foreman-Mackey} D.,  {Greengard} L.,  {Hogg} D.~W.,
  {O'Neil} M.,  2014, ArXiv e-prints arXiv:1403.6015

\bibitem[\protect\citeauthoryear{{Anantharamaiah}, {Viallefond}, {Mohan},
  {Goss} \& {Zhao}}{{Anantharamaiah} et~al.}{2000}]{2000ApJ...537..613A}
{Anantharamaiah} K.~R.,  {Viallefond} F.,  {Mohan} N.~R.,  {Goss} W.~M.,
  {Zhao} J.~H.,  2000, \apj, 537, 613

\bibitem[\protect\citeauthoryear{{Baars}, {Genzel}, {Pauliny-Toth} \&
  {Witzel}}{{Baars} et~al.}{1977}]{1977A&A....61...99B}
{Baars} J.~W.~M.,  {Genzel} R.,  {Pauliny-Toth} I.~I.~K.,    {Witzel} A.,
  1977, \aap, 61, 99

\bibitem[\protect\citeauthoryear{{Bell}}{{Bell}}{2003}]{2003ApJ...586..794B}
{Bell} E.~F.,  2003, \apj, 586, 794

\bibitem[\protect\citeauthoryear{{Borne}, {Bushouse}, {Lucas} \&
  {Colina}}{{Borne} et~al.}{2000}]{2000ApJ...529L..77B}
{Borne} K.~D.,  {Bushouse} H.,  {Lucas} R.~A.,    {Colina} L.,  2000, \apjl,
  529, L77

\bibitem[\protect\citeauthoryear{{Callingham}, {Gaensler}, {Ekers} \& {et.
  al.}}{{Callingham} et~al.}{2015}]{2015ApJ...809..168C}
{Callingham} J.~R.,  {Gaensler} B.~M.,  {Ekers} R.~D.,    {et. al.} 2015, \apj,
  809, 168

\bibitem[\protect\citeauthoryear{{Clemens}, {Scaife}, {Vega} \&
  {Bressan}}{{Clemens} et~al.}{2010}]{2010MNRAS.405..887C}
{Clemens} M.~S.,  {Scaife} A.,  {Vega} O.,    {Bressan} A.,  2010, \mnras, 405,
  887

\bibitem[\protect\citeauthoryear{{Clemens}, {Vega}, {Bressan}, {Granato},
  {Silva} \& {Panuzzo}}{{Clemens} et~al.}{2008}]{2008A&A...477...95C}
{Clemens} M.~S.,  {Vega} O.,  {Bressan} A.,  {Granato} G.~L.,  {Silva} L.,
  {Panuzzo} P.,  2008, \aap, 477, 95

\bibitem[\protect\citeauthoryear{{Condon}}{{Condon}}{1992}]{1992ARA&A..30..575C}
{Condon} J.~J.,  1992, \araa, 30, 575

\bibitem[\protect\citeauthoryear{{Condon}, {Cotton}, {Greisen}, {Yin} \& {et
  al.}}{{Condon} et~al.}{1998}]{1998AJ....115.1693C}
{Condon} J.~J.,  {Cotton} W.~D.,  {Greisen} E.~W.,  {Yin} Q.~F.,    {et al.}
  1998, \aj, 115, 1693

\bibitem[\protect\citeauthoryear{{Condon} \& {Yin}}{{Condon} \&
  {Yin}}{1990}]{1990ApJ...357...97C}
{Condon} J.~J.,  {Yin} Q.~F.,  1990, \apj, 357, 97

\bibitem[\protect\citeauthoryear{{de Zotti}, {Massardi}, {Negrello} \&
  {Wall}}{{de Zotti} et~al.}{2010}]{2010A&ARv..18....1D}
{de Zotti} G.,  {Massardi} M.,  {Negrello} M.,    {Wall} J.,  2010, \aapr, 18,
  1

\bibitem[\protect\citeauthoryear{{Dopita}, {Hart}, {McGregor}, {Oates},
  {Bloxham} \& {Jones}}{{Dopita} et~al.}{2007}]{2007Ap&SS.310..255D}
{Dopita} M.,  {Hart} J.,  {McGregor} P.,  {Oates} P.,  {Bloxham} G.,    {Jones}
  D.,  2007, \apss, 310, 255

\bibitem[\protect\citeauthoryear{{Dopita}, {Rhee}, {Farage}, {McGregor} \& {et
  al.}}{{Dopita} et~al.}{2010}]{2010Ap&SS.327..245D}
{Dopita} M.,  {Rhee} J.,  {Farage} C.,  {McGregor} P.,    {et al.} 2010, \apss,
  327, 245

\bibitem[\protect\citeauthoryear{{Downes} \& {Solomon}}{{Downes} \&
  {Solomon}}{1998}]{1998ApJ...507..615D}
{Downes} D.,  {Solomon} P.~M.,  1998, \apj, 507, 615

\bibitem[\protect\citeauthoryear{{Duc}, {Mirabel} \& {Maza}}{{Duc}
  et~al.}{1997}]{1997A&AS..124..533D}
{Duc} P.-A.,  {Mirabel} I.~F.,    {Maza} J.,  1997, \aaps, 124

\bibitem[\protect\citeauthoryear{{Engel}, {Davies}, {Genzel}, {Tacconi},
  {Sturm} \& {Downes}}{{Engel} et~al.}{2011}]{2011ApJ...729...58E}
{Engel} H.,  {Davies} R.~I.,  {Genzel} R.,  {Tacconi} L.~J.,  {Sturm} E.,
  {Downes} D.,  2011, \apj, 729, 58

\bibitem[\protect\citeauthoryear{{Farrah}, {Afonso}, {Efstathiou},
  {Rowan-Robinson}, {Fox} \& {Clements}}{{Farrah}
  et~al.}{2003}]{2003MNRAS.343..585F}
{Farrah} D.,  {Afonso} J.,  {Efstathiou} A.,  {Rowan-Robinson} M.,  {Fox} M.,
   {Clements} D.,  2003, \mnras, 343, 585

\bibitem[\protect\citeauthoryear{{Farrah}, {Rowan-Robinson}, {Oliver},
  {Serjeant}, {Borne}, {Lawrence}, {Lucas}, {Bushouse} \& {Colina}}{{Farrah}
  et~al.}{2001}]{2001MNRAS.326.1333F}
{Farrah} D.,  {Rowan-Robinson} M.,  {Oliver} S.,  {Serjeant} S.,  {Borne} K.,
  {Lawrence} A.,  {Lucas} R.~A.,  {Bushouse} H.,    {Colina} L.,  2001, \mnras,
  326, 1333

\bibitem[\protect\citeauthoryear{{Feroz}, {Hobson} \& {Bridges}}{{Feroz}
  et~al.}{2009}]{2009MNRAS.398.1601F}
{Feroz} F.,  {Hobson} M.~P.,    {Bridges} M.,  2009, \mnras, 398, 1601

\bibitem[\protect\citeauthoryear{{Foreman-Mackey}, {Hogg}, {Lang} \&
  {Goodman}}{{Foreman-Mackey} et~al.}{2013}]{2013PASP..125..306F}
{Foreman-Mackey} D.,  {Hogg} D.~W.,  {Lang} D.,    {Goodman} J.,  2013, \pasp,
  125, 306

\bibitem[\protect\citeauthoryear{{Franzen}, {Jackson}, {Offringa} \& {et
  al.}}{{Franzen} et~al.}{2016}]{2016MNRAS.459.3314F}
{Franzen} T.~M.~O.,  {Jackson} C.~A.,  {Offringa} A.~R.,    {et al.} 2016,
  \mnras, 459, 3314

\bibitem[\protect\citeauthoryear{{Frater}, {Brooks} \& {Whiteoak}}{{Frater}
  et~al.}{1992}]{1992JEEEA..12..103F}
{Frater} R.~H.,  {Brooks} J.~W.,    {Whiteoak} J.~B.,  1992, Journal of
  Electrical and Electronics Engineering Australia, 12, 103

\bibitem[\protect\citeauthoryear{{Galvin}, {Seymour}, {Filipovi{\'c}},
  {Tothill}, {Marvil}, {Drouart}, {Symeonidis} \& {Huynh}}{{Galvin}
  et~al.}{2016}]{2016MNRAS.461..825G}
{Galvin} T.~J.,  {Seymour} N.,  {Filipovi{\'c}} M.~D.,  {Tothill} N.~F.~H.,
  {Marvil} J.,  {Drouart} G.,  {Symeonidis} M.,    {Huynh} M.~T.,  2016,
  \mnras, 461, 825

\bibitem[\protect\citeauthoryear{{Gooch}}{{Gooch}}{1996}]{1996ASPC..101...80G}
{Gooch} R.,  1996, in {Jacoby} G.~H.,  {Barnes} J.,  eds, Astronomical Data
  Analysis Software and Systems V Vol.~101 of Astronomical Society of the
  Pacific Conference Series, {Karma: a Visualization Test-Bed}.
p.~80

\bibitem[\protect\citeauthoryear{{Goodman} \& {Weare}}{{Goodman} \&
  {Weare}}{2010}]{GW}
{Goodman} J.,  {Weare} J.,  2010, {Commun. Appl. Math. Comput. Sci.}, 5, 65

\bibitem[\protect\citeauthoryear{{Gregory}, {Vavasour}, {Scott} \&
  {Condon}}{{Gregory} et~al.}{1994}]{1994ApJS...90..173G}
{Gregory} P.~C.,  {Vavasour} J.~D.,  {Scott} W.~K.,    {Condon} J.~J.,  1994,
  \apjs, 90, 173

\bibitem[\protect\citeauthoryear{{Gruppioni}, {Pozzi}, {Zamorani}, {Ciliegi},
  {Lari}, {Calabrese}, {La Franca} \& {Matute}}{{Gruppioni}
  et~al.}{2003}]{2003MNRAS.341L...1G}
{Gruppioni} C.,  {Pozzi} F.,  {Zamorani} G.,  {Ciliegi} P.,  {Lari} C.,
  {Calabrese} E.,  {La Franca} F.,    {Matute} I.,  2003, \mnras, 341, L1

\bibitem[\protect\citeauthoryear{{Hancock}, {Murphy}, {Gaensler}, {Hopkins} \&
  {Curran}}{{Hancock} et~al.}{2012}]{2012MNRAS.422.1812H}
{Hancock} P.~J.,  {Murphy} T.,  {Gaensler} B.~M.,  {Hopkins} A.,    {Curran}
  J.~R.,  2012, \mnras, 422, 1812

\bibitem[\protect\citeauthoryear{{Helou}, {Khan}, {Malek} \& {Boehmer}}{{Helou}
  et~al.}{1988}]{1988ApJS...68..151H}
{Helou} G.,  {Khan} I.~R.,  {Malek} L.,    {Boehmer} L.,  1988, \apjs, 68, 151

\bibitem[\protect\citeauthoryear{{Helou}, {Soifer} \& {Rowan-Robinson}}{{Helou}
  et~al.}{1985}]{1985ApJ...298L...7H}
{Helou} G.,  {Soifer} B.~T.,    {Rowan-Robinson} M.,  1985, \apjl, 298, L7

\bibitem[\protect\citeauthoryear{{Hildebrand}}{{Hildebrand}}{1983}]{1983QJRAS..24..267H}
{Hildebrand} R.~H.,  1983, \qjras, 24, 267

\bibitem[\protect\citeauthoryear{{Hughes}, {Staveley-Smith}, {Kim}, {Wolleben}
  \& {Filipovi{\'c}}}{{Hughes} et~al.}{2007}]{2007MNRAS.382..543H}
{Hughes} A.,  {Staveley-Smith} L.,  {Kim} S.,  {Wolleben} M.,
  {Filipovi{\'c}} M.,  2007, \mnras, 382, 543

\bibitem[\protect\citeauthoryear{{Hughes}, {Wong} \& {et al.}}{{Hughes}
  et~al.}{2006}]{2006MNRAS.370..363H}
{Hughes} A.,  {Wong} T.,    {et al.} 2006, \mnras, 370, 363

\bibitem[\protect\citeauthoryear{{Hurley-Walker}, {Callingham}, {Hancock} \&
  {et. al.}}{{Hurley-Walker} et~al.}{2017}]{2017MNRAS.464.1146H}
{Hurley-Walker} N.,  {Callingham} J.~R.,  {Hancock} P.~J.,    {et. al.} 2017,
  \mnras, 464, 1146

\bibitem[\protect\citeauthoryear{{Huynh}, {Hopkins}, {Lenc}, {Mao} \& {et.
  al.}}{{Huynh} et~al.}{2012}]{2012MNRAS.426.2342H}
{Huynh} M.~T.,  {Hopkins} A.~M.,  {Lenc} E.,  {Mao} M.~Y.,    {et. al.} 2012,
  \mnras, 426, 2342

\bibitem[\protect\citeauthoryear{{Ibar}, {Ivison}, {Biggs}, {Lal}, {Best} \&
  {Green}}{{Ibar} et~al.}{2009}]{2009MNRAS.397..281I}
{Ibar} E.,  {Ivison} R.~J.,  {Biggs} A.~D.,  {Lal} D.~V.,  {Best} P.~N.,
  {Green} D.~A.,  2009, \mnras, 397, 281

\bibitem[\protect\citeauthoryear{{Ivison}, {Magnelli} \& {et al.}}{{Ivison}
  et~al.}{2010}]{2010A&A...518L..31I}
{Ivison} R.~J.,  {Magnelli} B.,    {et al.} 2010, \aap, 518, L31

\bibitem[\protect\citeauthoryear{{Jarvis}, {Seymour} \& {et. al.}}{{Jarvis}
  et~al.}{2015}]{2015aska.confE..68J}
{Jarvis} M.,  {Seymour} N.,    {et. al.} 2015, Advancing Astrophysics with the
  Square Kilometre Array (AASKA14), p.~68

\bibitem[\protect\citeauthoryear{{Kapinska}, {Staveley-Smith}, {Crocker} \& {et
  al.}}{{Kapinska} et~al.}{2017}]{2017arXiv170202434K}
{Kapinska} A.~D.,  {Staveley-Smith} L.,  {Crocker} R.,    {et al.} 2017, ArXiv
  e-prints arXiv:1702.02434, ApJ, in press

\bibitem[\protect\citeauthoryear{Kass \& Raftery}{Kass \&
  Raftery}{1995}]{kass1995bayes}
Kass R.~E.,  Raftery A.~E.,  1995, Journal of the american statistical
  association, 90, 773

\bibitem[\protect\citeauthoryear{{Klaas}, {Haas} \& {et al.}}{{Klaas}
  et~al.}{2001}]{2001A&A...379..823K}
{Klaas} U.,  {Haas} M.,    {et al.} 2001, \aap, 379, 823

\bibitem[\protect\citeauthoryear{{Komatsu}, {Dunkley}, {Nolta}, {Bennett} \&
  {et al.}}{{Komatsu} et~al.}{2009}]{2009ApJS..180..330K}
{Komatsu} E.,  {Dunkley} J.,  {Nolta} M.~R.,  {Bennett} C.~L.,    {et al.}
  2009, \apjs, 180, 330

\bibitem[\protect\citeauthoryear{{Lacki} \& {Thompson}}{{Lacki} \&
  {Thompson}}{2010}]{2010ApJ...717..196L}
{Lacki} B.~C.,  {Thompson} T.~A.,  2010, \apj, 717, 196

\bibitem[\protect\citeauthoryear{{Large}, {Cram} \& {Burgess}}{{Large}
  et~al.}{1991}]{1991Obs...111...72L}
{Large} M.~I.,  {Cram} L.~E.,    {Burgess} A.~M.,  1991, The Observatory, 111,
  72

\bibitem[\protect\citeauthoryear{{Large}, {Mills}, {Little}, {Crawford} \&
  {Sutton}}{{Large} et~al.}{1981}]{1981MNRAS.194..693L}
{Large} M.~I.,  {Mills} B.~Y.,  {Little} A.~G.,  {Crawford} D.~F.,    {Sutton}
  J.~M.,  1981, \mnras, 194, 693

\bibitem[\protect\citeauthoryear{{Lenc} \& {Tingay}}{{Lenc} \&
  {Tingay}}{2009}]{2009AJ....137..537L}
{Lenc} E.,  {Tingay} S.~J.,  2009, \aj, 137, 537

\bibitem[\protect\citeauthoryear{{Leroy}, {Evans}, {Momjian}, {Murphy} \& {et
  al.}}{{Leroy} et~al.}{2011}]{2011ApJ...739L..25L}
{Leroy} A.~K.,  {Evans} A.~S.,  {Momjian} E.,  {Murphy} E.,    {et al.} 2011,
  \apjl, 739, L25

\bibitem[\protect\citeauthoryear{{Lonsdale}, {Cappallo}, {Morales}, {Briggs} \&
  {et al.}}{{Lonsdale} et~al.}{2009}]{2009IEEEP..97.1497L}
{Lonsdale} C.~J.,  {Cappallo} R.~J.,  {Morales} M.~F.,  {Briggs} F.~H.,    {et
  al.} 2009, IEEE Proceedings, 97, 1497

\bibitem[\protect\citeauthoryear{{Mao}, {Huynh}, {Norris}, {Dickinson},
  {Frayer}, {Helou} \& {Monkiewicz}}{{Mao} et~al.}{2011}]{2011ApJ...731...79M}
{Mao} M.~Y.,  {Huynh} M.~T.,  {Norris} R.~P.,  {Dickinson} M.,  {Frayer} D.,
  {Helou} G.,    {Monkiewicz} J.~A.,  2011, \apj, 731, 79

\bibitem[\protect\citeauthoryear{{Mauch}, {Murphy}, {Buttery}, {Curran} \& {et
  al.}}{{Mauch} et~al.}{2003}]{2003MNRAS.342.1117M}
{Mauch} T.,  {Murphy} T.,  {Buttery} H.~J.,  {Curran} J.,    {et al.} 2003,
  \mnras, 342, 1117

\bibitem[\protect\citeauthoryear{{Mauch}, {Murphy} \& {et al.}}{{Mauch}
  et~al.}{2013}]{2013yCat.8081....0M}
{Mauch} T.,  {Murphy} T.,    {et al.} 2013, VizieR Online Data Catalog, 8081, 0

\bibitem[\protect\citeauthoryear{{Middelberg}, {Sault} \&
  {Kesteven}}{{Middelberg} et~al.}{2006}]{2006PASA...23..147M}
{Middelberg} E.,  {Sault} R.~J.,    {Kesteven} M.~J.,  2006, \pasa, 23, 147

\bibitem[\protect\citeauthoryear{{Mooley}, {Hallinan}, {Bourke}, {Horesh} \&
  {et al.}}{{Mooley} et~al.}{2016}]{2016ApJ...818..105M}
{Mooley} K.~P.,  {Hallinan} G.,  {Bourke} S.,  {Horesh} A.,    {et al.} 2016,
  \apj, 818, 105

\bibitem[\protect\citeauthoryear{{Moshir}}{{Moshir}}{1990}]{1990IRASF.C......0M}
{Moshir} M.,  1990, in IRAS Faint Source Catalogue, version 2.0 (1990) {IRAS
  Faint Source Catalogue, version 2.0.}.
p.~0

\bibitem[\protect\citeauthoryear{{Murakami}, {Baba}, {Barthel}, {Clements} \&
  {et al.}}{{Murakami} et~al.}{2007}]{2007PASJ...59S.369M}
{Murakami} H.,  {Baba} H.,  {Barthel} P.,  {Clements} D.~L.,    {et al.} 2007,
  \pasj, 59, S369

\bibitem[\protect\citeauthoryear{{Murphy}}{{Murphy}}{2009}]{2009ApJ...706..482M}
{Murphy} E.~J.,  2009, \apj, 706, 482

\bibitem[\protect\citeauthoryear{{Murphy}}{{Murphy}}{2013}]{2013ApJ...777...58M}
{Murphy} E.~J.,  2013, \apj, 777, 58

\bibitem[\protect\citeauthoryear{{Murphy}, {Bremseth}, {Mason}, {Condon},
  {Schinnerer}, {Aniano}, {Armus}, {Helou}, {Turner} \& {Jarrett}}{{Murphy}
  et~al.}{2012}]{2012ApJ...761...97M}
{Murphy} E.~J.,  {Bremseth} J.,  {Mason} B.~S.,  {Condon} J.~J.,  {Schinnerer}
  E.,  {Aniano} G.,  {Armus} L.,  {Helou} G.,  {Turner} J.~L.,    {Jarrett}
  T.~H.,  2012, \apj, 761, 97

\bibitem[\protect\citeauthoryear{{Murphy}, {Helou}, {Condon}, {Schinnerer},
  {Turner}, {Beck}, {Mason}, {Chary} \& {Armus}}{{Murphy}
  et~al.}{2010}]{2010ApJ...709L.108M}
{Murphy} E.~J.,  {Helou} G.,  {Condon} J.~J.,  {Schinnerer} E.,  {Turner}
  J.~L.,  {Beck} R.,  {Mason} B.~S.,  {Chary} R.-R.,    {Armus} L.,  2010,
  \apjl, 709, L108

\bibitem[\protect\citeauthoryear{{Murphy}, {Helou} \& {et al.}}{{Murphy}
  et~al.}{2006}]{2006ApJ...651L.111M}
{Murphy} E.~J.,  {Helou} G.,    {et al.} 2006, \apjl, 651, L111

\bibitem[\protect\citeauthoryear{{Neff} \& {Ulvestad}}{{Neff} \&
  {Ulvestad}}{2000}]{2000AJ....120..670N}
{Neff} S.~G.,  {Ulvestad} J.~S.,  2000, \aj, 120, 670

\bibitem[\protect\citeauthoryear{{Niklas}, {Klein} \& {Wielebinski}}{{Niklas}
  et~al.}{1997}]{1997A&A...322...19N}
{Niklas} S.,  {Klein} U.,    {Wielebinski} R.,  1997, \aap, 322, 19

\bibitem[\protect\citeauthoryear{{Norris}, {Hopkins} \& {et. al.}}{{Norris}
  et~al.}{2011}]{2011PASA...28..215N}
{Norris} R.~P.,  {Hopkins} A.~M.,    {et. al.} 2011, \pasa, 28, 215

\bibitem[\protect\citeauthoryear{Partridge, L{\'o}pez-Caniego, Perley, Stevens,
  Butler, Rocha, Walter \& Zacchei}{Partridge
  et~al.}{2016}]{0004-637X-821-1-61}
Partridge B.,  L{\'o}pez-Caniego M.,  Perley R.~A.,  Stevens J.,  Butler B.~J.,
   Rocha G.,  Walter B.,    Zacchei A.,  2016, The Astrophysical Journal, 821,
  61

\bibitem[\protect\citeauthoryear{{Paturel}, {Petit}, {Prugniel}, {Theureau},
  {Rousseau}, {Brouty}, {Dubois} \& {Cambr{\'e}sy}}{{Paturel}
  et~al.}{2003}]{2003A&A...412...45P}
{Paturel} G.,  {Petit} C.,  {Prugniel} P.,  {Theureau} G.,  {Rousseau} J.,
  {Brouty} M.,  {Dubois} P.,    {Cambr{\'e}sy} L.,  2003, \aap, 412, 45

\bibitem[\protect\citeauthoryear{{Prandoni} \& {Seymour}}{{Prandoni} \&
  {Seymour}}{2015}]{2015aska.confE..67P}
{Prandoni} I.,  {Seymour} N.,  2015, Advancing Astrophysics with the Square
  Kilometre Array (AASKA14), p.~67

\bibitem[\protect\citeauthoryear{{Price} \& {Duric}}{{Price} \&
  {Duric}}{1992}]{1992ApJ...401...81P}
{Price} R.,  {Duric} N.,  1992, \apj, 401, 81

\bibitem[\protect\citeauthoryear{{Rampadarath}, {Morgan}, {Lenc} \&
  {Tingay}}{{Rampadarath} et~al.}{2014}]{2014AJ....147....5R}
{Rampadarath} H.,  {Morgan} J.~S.,  {Lenc} E.,    {Tingay} S.~J.,  2014, \aj,
  147, 5

\bibitem[\protect\citeauthoryear{{Rasmussen} \& {Williams}}{{Rasmussen} \&
  {Williams}}{2006}]{gaussianprocess}
{Rasmussen} C.~E.,  {Williams} C.~K.~I.,  2006, Gaussian Processes for Machine
  Learning.
MIT Press, Cambridge, MA, p.~248

\bibitem[\protect\citeauthoryear{Reynolds}{Reynolds}{1994}]{1934}
Reynolds J.,  1994, Technical Report AT/39.3/040, {A revised flux scale for the
  AT compact array}.
ATNF Memo

\bibitem[\protect\citeauthoryear{{Sault}, {Teuben} \& {Wright}}{{Sault}
  et~al.}{1995}]{1995ASPC...77..433S}
{Sault} R.~J.,  {Teuben} P.~J.,    {Wright} M.~C.~H.,  1995, in {Shaw} R.~A.,
  {Payne} H.~E.,   {Hayes} J.~J.~E.,  eds, Astronomical Data Analysis Software
  and Systems IV Vol.~77 of Astronomical Society of the Pacific Conference
  Series, {A Retrospective View of MIRIAD}.
p.~433

\bibitem[\protect\citeauthoryear{{Sault} \& {Wieringa}}{{Sault} \&
  {Wieringa}}{1994}]{1994A&AS..108..585S}
{Sault} R.~J.,  {Wieringa} M.~H.,  1994, \aaps, 108, 585

\bibitem[\protect\citeauthoryear{{Saunders}, {Sutherland}, {Maddox}, {Keeble},
  {Oliver} \& {et al.}}{{Saunders} et~al.}{2000}]{2000MNRAS.317...55S}
{Saunders} W.,  {Sutherland} W.~J.,  {Maddox} S.~J.,  {Keeble} O.,  {Oliver}
  S.~J.,    {et al.} 2000, \mnras, 317, 55

\bibitem[\protect\citeauthoryear{{Seymour}, {Dwelly} \& {et al.}}{{Seymour}
  et~al.}{2008}]{2008MNRAS.386.1695S}
{Seymour} N.,  {Dwelly} T.,    {et al.} 2008, \mnras, 386, 1695

\bibitem[\protect\citeauthoryear{{Smith}, {Hardcastle}, {Jarvis}, {Maddox} \&
  {et al.}}{{Smith} et~al.}{2013}]{2013MNRAS.436.2435S}
{Smith} D.~J.~B.,  {Hardcastle} M.~J.,  {Jarvis} M.~J.,  {Maddox} S.~J.,    {et
  al.} 2013, \mnras, 436, 2435

\bibitem[\protect\citeauthoryear{{Smith}, {Lonsdale}, {Lonsdale} \&
  {Diamond}}{{Smith} et~al.}{1998}]{1998ApJ...493L..17S}
{Smith} H.~E.,  {Lonsdale} C.~J.,  {Lonsdale} C.~J.,    {Diamond} P.~J.,  1998,
  \apjl, 493, L17

\bibitem[\protect\citeauthoryear{{Tingay}, {Goeke}, {Bowman}, {Emrich} \& {et
  al.}}{{Tingay} et~al.}{2013}]{2013PASA...30....7T}
{Tingay} S.~J.,  {Goeke} R.,  {Bowman} J.~D.,  {Emrich} D.,    {et al.} 2013,
  \pasa, 30, e007

\bibitem[\protect\citeauthoryear{{Waldram}, {Bolton}, {Pooley} \&
  {Riley}}{{Waldram} et~al.}{2007}]{2007MNRAS.379.1442W}
{Waldram} E.~M.,  {Bolton} R.~C.,  {Pooley} G.~G.,    {Riley} J.~M.,  2007,
  \mnras, 379, 1442

\bibitem[\protect\citeauthoryear{{Wang}, {Rowan-Robinson}, {Norberg}, {Heinis}
  \& {Han}}{{Wang} et~al.}{2014}]{2014MNRAS.442.2739W}
{Wang} L.,  {Rowan-Robinson} M.,  {Norberg} P.,  {Heinis} S.,    {Han} J.,
  2014, \mnras, 442, 2739

\bibitem[\protect\citeauthoryear{{Wang} \& {Fields}}{{Wang} \&
  {Fields}}{2016}]{2016arXiv161207290W}
{Wang} X.,  {Fields} B.~D.,  2016, ArXiv e-prints arXiv:1612.07290

\bibitem[\protect\citeauthoryear{{Wayth}, {Lenc}, {Bell} \& {et}}{{Wayth}
  et~al.}{2015}]{2015PASA...32...25W}
{Wayth} R.~B.,  {Lenc} E.,  {Bell} M.~E.,    {et} a.,  2015, \pasa, 32, e025

\bibitem[\protect\citeauthoryear{{Wills}, {Pedlar}, {Muxlow} \&
  {Wilkinson}}{{Wills} et~al.}{1997}]{1997MNRAS.291..517W}
{Wills} K.~A.,  {Pedlar} A.,  {Muxlow} T.~W.~B.,    {Wilkinson} P.~N.,  1997,
  \mnras, 291, 517

\bibitem[\protect\citeauthoryear{{Wilman}, {Miller}, {Jarvis}, {Mauch},
  {Levrier}, {Abdalla}, {Rawlings}, {Kl{\"o}ckner}, {Obreschkow}, {Olteanu} \&
  {Young}}{{Wilman} et~al.}{2008}]{2008MNRAS.388.1335W}
{Wilman} R.~J.,  {Miller} L.,  {Jarvis} M.~J.,  {Mauch} T.,  {Levrier} F.,
  {Abdalla} F.~B.,  {Rawlings} S.,  {Kl{\"o}ckner} H.-R.,  {Obreschkow} D.,
  {Olteanu} D.,    {Young} S.,  2008, \mnras, 388, 1335

\bibitem[\protect\citeauthoryear{{Wilson}, {Ferris} \& {et al.}}{{Wilson}
  et~al.}{2011}]{2011MNRAS.416..832W}
{Wilson} W.~E.,  {Ferris} R.~H.,    {et al.} 2011, \mnras, 416, 832

\bibitem[\protect\citeauthoryear{{Yamamura}, {Makiuti}, {Ikeda}, {Fukuda} \&
  {et al.}}{{Yamamura} et~al.}{2010}]{2010yCat.2298....0Y}
{Yamamura} I.,  {Makiuti} S.,  {Ikeda} N.,  {Fukuda} Y.,    {et al.} 2010,
  VizieR Online Data Catalog, 2298

\bibitem[\protect\citeauthoryear{{Yun}, {Reddy} \& {Condon}}{{Yun}
  et~al.}{2001}]{2001ApJ...554..803Y}
{Yun} M.~S.,  {Reddy} N.~A.,    {Condon} J.~J.,  2001, \apj, 554, 803

\end{thebibliography}

\onecolumn
\begin{landscape}

\appendix

\section{Flux Density Measurements}
In this section we list the flux density measurements used throughout the modelling described in Section 2 and 3. For measurements acquired through the NED database we include references to the source of the measurements.

\begin{table}
\centering
\caption{An overview of all radio-continuum measurements obtained from the MWA GLEAM project \citep{2015PASA...32...25W}. Sources with upwards of 20 measurements were taken directly from the MWA GLEAM catalogue \citep{2017MNRAS.464.1146H}. Otherwise, sources with only four measurements at frequencies of 88, 119, 155 and 200\,MHz were obtained using the \textsc{priorised} option available in the \textsc{aegean} packages.  \label{table:mwa_measurements}}
\tiny
\begin{tabular}{lccccccccccccccccccccccccc}\toprule
Source & \multicolumn{12}{c}{MWA GLEAM}\\
 \textit{IRAS} & 76\,MHz & 84\,MHz & 88\,MHz & 92\,MHz & 99\,MHz & 107\,MHz & 115\,MHz & 119\,MHz & 122\,MHz & 130\,MHz & 143\,MHz & 151\,MHz \\
  & (mJy) & (mJy) & (mJy) & (mJy) & (mJy) & (mJy) & (mJy) & (mJy) & (mJy) & (mJy) & (mJy) & (mJy) \\
\midrule
F00198-7926 & 47.9$\pm$121.6 & -- & -- &237.4$\pm$97.8 &226.4$\pm$106.4 &101.2$\pm$63.3 &5.7$\pm$48.1 & -- &179.9$\pm$50.0 &218.9$\pm$47.0 &39.5$\pm$34.0 &156.4$\pm$39.6 &\\
\hline
F00199-7426 &  -- & -- &141.0$\pm$53.9 & -- & -- & -- & -- &25.6$\pm$25.6 & -- & -- & -- & -- &\\
\hline
F01268-5436 & 361.9$\pm$81.2 &237.9$\pm$60.0 & -- &83.7$\pm$48.9 &28.0$\pm$48.6 &87.7$\pm$38.3 &122.5$\pm$33.4 & -- &123.9$\pm$30.2 &135.5$\pm$29.4 &147.5$\pm$30.0 &112.1$\pm$25.7 &\\
\hline
F01388-4618 &  -- & -- &32.4$\pm$32.4 & -- & -- & -- & -- &15.2$\pm$15.2 & -- & -- & -- & -- &\\
\hline
F01419-6826 &  -- & -- &31.0$\pm$31.0 & -- & -- & -- & -- &36.6$\pm$14.5 & -- & -- & -- & -- &\\
\hline
F02364-4751 & 32.5$\pm$73.8 &139.6$\pm$54.3 & -- &73.6$\pm$45.5 &27.3$\pm$43.3 &2.3$\pm$35.7 &68.0$\pm$29.6 & -- &98.8$\pm$27.7 &69.2$\pm$24.7 &52.4$\pm$17.5 &41.4$\pm$15.4 &\\
\hline
F03068-5346 &  -- &170.0$\pm$52.5 & -- &46.3$\pm$45.3 &77.9$\pm$44.0 &98.3$\pm$33.3 &91.4$\pm$28.3 & -- &74.1$\pm$25.2 &83.2$\pm$24.3 &80.2$\pm$21.5 &38.9$\pm$18.7 &\\
\hline
F03481-4012 &  -- &80.4$\pm$59.0 & -- &93.2$\pm$52.3 &114.5$\pm$48.6 &138.7$\pm$37.3 &69.4$\pm$29.9 & -- &75.0$\pm$28.2 &64.0$\pm$25.0 &76.0$\pm$24.8 &66.4$\pm$22.2 &\\
\hline
F04063-3236 & 195.1$\pm$76.5 & -- & -- & -- & -- &54.2$\pm$37.3 &34.9$\pm$30.1 & -- &29.4$\pm$28.1 &20.7$\pm$25.6 &18.3$\pm$21.3 &44.4$\pm$19.0 &\\
\hline
F06021-4509 &  -- & -- &38.7$\pm$38.7 & -- & -- & -- & -- &20.9$\pm$20.9 & -- & -- & -- & -- &\\
\hline
F06035-7102 & 898.0$\pm$144.6 &568.7$\pm$97.2 & -- &686.1$\pm$101.7 &712.2$\pm$106.6 &634.5$\pm$81.8 &530.8$\pm$67.7 & -- &536.6$\pm$65.2 &510.9$\pm$62.0 &474.8$\pm$56.0 &465.7$\pm$54.3 &\\
\hline
F06206-6315 &  -- & -- &30.8$\pm$30.8 & -- & -- & -- & -- &14.3$\pm$14.3 & -- & -- & -- & -- &\\
\hline
F18582-5558 &  -- & -- &40.4$\pm$40.4 & -- & -- & -- & -- &21.8$\pm$21.8 & -- & -- & -- & -- &\\
\hline
F20117-3249 &  -- & -- &201.0$\pm$47.8 & -- & -- & -- & -- &131.0$\pm$39.2 & -- & -- & -- & -- &\\
\hline
F20445-6218 &  -- & -- &43.7$\pm$43.7 & -- & -- & -- & -- &20.0$\pm$20.0 & -- & -- & -- & -- &\\
\hline
F21178-6349 &  -- & -- &47.9$\pm$47.9 & -- & -- & -- & -- &20.6$\pm$20.6 & -- & -- & -- & -- &\\
\hline
F21292-4953 & 40.8$\pm$80.3 &140.7$\pm$65.0 & -- &50.5$\pm$56.8 &51.1$\pm$47.1 &135.1$\pm$39.2 &61.0$\pm$29.8 & -- &96.4$\pm$26.8 &79.9$\pm$25.4 &71.2$\pm$19.3 &51.1$\pm$16.5 &\\
\hline
F21295-4634 &  -- & -- &28.7$\pm$28.7 & -- & -- & -- & -- &17.4$\pm$17.4 & -- & -- & -- & -- &\\
\hline
F23389-6139 & 114.8$\pm$62.1 &140.2$\pm$50.7 & -- &123.3$\pm$48.0 &145.2$\pm$43.6 &191.0$\pm$39.4 &202.2$\pm$34.9 & -- &181.1$\pm$35.5 &229.3$\pm$33.5 &298.3$\pm$31.4 &268.4$\pm$28.8 &\\
\hline
\toprule
Source & \multicolumn{12}{c}{MWA GLEAM}\\
 \textit{IRAS} & 155\,MHz & 158\,MHz & 166\,MHz & 174\,MHz & 181\,MHz & 189\,MHz & 197\,MHz & 200\,MHz & 204\,MHz & 212\,MHz & 220\,MHz & 227\,MHz \\
  & (mJy) & (mJy) & (mJy) & (mJy) & (mJy) & (mJy) & (mJy) & (mJy) & (mJy) & (mJy) & (mJy) & (mJy) \\
\midrule
F00198-7926 &  -- &149.7$\pm$38.3 &68.6$\pm$34.1 &188.8$\pm$42.2 &135.1$\pm$40.5 &81.0$\pm$36.9 &156.3$\pm$39.2 & -- &108.2$\pm$38.5 &114.5$\pm$40.4 &137.0$\pm$41.5 & -- &\\
\hline
F00199-7426 & 67.5$\pm$19.1 & -- & -- & -- & -- & -- & -- &37.4$\pm$12.0 & -- & -- & -- & -- &\\
\hline
F01268-5436 &  -- &118.6$\pm$23.8 &65.8$\pm$22.0 &88.1$\pm$22.7 &79.0$\pm$22.2 &102.9$\pm$22.0 &85.1$\pm$23.1 & -- &97.4$\pm$19.1 &51.6$\pm$18.2 &58.5$\pm$18.0 &86.6$\pm$19.6 &\\
\hline
F01388-4618 & 20.1$\pm$7.4 & -- & -- & -- & -- & -- & -- &25.6$\pm$5.7 & -- & -- & -- & -- &\\
\hline
F01419-6826 & 9.8$\pm$9.8 & -- & -- & -- & -- & -- & -- &19.3$\pm$19.3 & -- & -- & -- & -- &\\
\hline
F02364-4751 &  -- &58.5$\pm$13.8 &52.8$\pm$13.2 &61.2$\pm$14.7 &27.0$\pm$12.2 &57.5$\pm$12.7 &61.1$\pm$12.6 & -- &40.9$\pm$14.5 &21.7$\pm$12.8 &61.7$\pm$13.7 &29.1$\pm$13.6 &\\
\hline
F03068-5346 &  -- &104.2$\pm$18.4 &107.4$\pm$18.6 &91.7$\pm$17.9 &74.0$\pm$15.7 &96.7$\pm$16.5 &77.0$\pm$16.7 & -- &74.3$\pm$14.3 &74.3$\pm$14.4 &49.7$\pm$12.3 &62.7$\pm$13.0 &\\
\hline
F03481-4012 &  -- &40.1$\pm$21.2 &59.4$\pm$20.2 &25.2$\pm$20.4 &84.4$\pm$21.4 &91.4$\pm$21.4 &70.6$\pm$19.4 & -- &57.0$\pm$15.1 &70.3$\pm$15.2 &62.2$\pm$15.1 &82.3$\pm$15.6 &\\
\hline
F04063-3236 &  -- &22.3$\pm$18.0 &40.8$\pm$16.2 &3.1$\pm$15.5 &31.0$\pm$13.3 &15.8$\pm$12.9 &40.5$\pm$12.2 & -- &47.0$\pm$14.2 &10.4$\pm$12.5 &47.9$\pm$13.3 &58.2$\pm$14.1 &\\
\hline
F06021-4509 & 48.8$\pm$14.2 & -- & -- & -- & -- & -- & -- &7.4$\pm$7.4 & -- & -- & -- & -- &\\
\hline
F06035-7102 &  -- &483.9$\pm$54.9 &480.6$\pm$55.2 &534.6$\pm$60.5 &458.3$\pm$54.4 &465.5$\pm$54.6 &375.6$\pm$48.2 & -- &436.4$\pm$52.7 &412.5$\pm$49.9 &366.9$\pm$45.4 &408.6$\pm$49.9 &\\
\hline
F06206-6315 & 30.8$\pm$8.8 & -- & -- & -- & -- & -- & -- &23.5$\pm$8.4 & -- & -- & -- & -- &\\
\hline
F18582-5558 & 21.9$\pm$21.9 & -- & -- & -- & -- & -- & -- &12.6$\pm$12.6 & -- & -- & -- & -- &\\
\hline
F20117-3249 & 22.2$\pm$22.2 & -- & -- & -- & -- & -- & -- &86.5$\pm$15.4 & -- & -- & -- & -- &\\
\hline
F20445-6218 & 16.8$\pm$16.8 & -- & -- & -- & -- & -- & -- &15.6$\pm$15.6 & -- & -- & -- & -- &\\
\hline
F21178-6349 & 14.2$\pm$14.2 & -- & -- & -- & -- & -- & -- &13.0$\pm$13.0 & -- & -- & -- & -- &\\
\hline
F21292-4953 &  -- &78.2$\pm$17.4 &47.9$\pm$14.8 &23.0$\pm$15.3 &55.5$\pm$16.2 &24.0$\pm$15.2 &82.0$\pm$16.7 & -- &72.7$\pm$17.4 &43.9$\pm$17.9 &28.6$\pm$16.6 &49.5$\pm$17.3 &\\
\hline
F21295-4634 & 10.5$\pm$10.5 & -- & -- & -- & -- & -- & -- &21.7$\pm$8.0 & -- & -- & -- & -- &\\
\hline
F23389-6139 &  -- &239.4$\pm$27.4 &273.0$\pm$27.5 &239.8$\pm$26.3 &286.6$\pm$31.7 &317.2$\pm$30.7 &292.4$\pm$29.5 & -- &272.5$\pm$29.2 &250.0$\pm$27.6 &269.6$\pm$26.6 &255.5$\pm$26.8 &\\
\hline
\end{tabular}
\end{table}

\tabcolsep=0.1cm
\begin{table}
\centering
\caption{An overview of all flux density measurements, obtained from the literature, that were used for sources in our samples. `--' denote sources without a measurement for that survey or instrument. References: SUMSS - \citet{2013yCat.8081....0M}, NVSS - \citet{1998AJ....115.1693C}, PMN - \citet{1994ApJS...90..173G}, \textit{IRAS} - \citet{1990IRASF.C......0M}, \textit{AKARI} - \citet{2010yCat.2298....0Y}, \textit{ISOPHOT} - \citet{2001A&A...379..823K}.  \label{table:ned_measurements}}
\tiny
\begin{tabular}{cccccccccccccccc}\toprule
Source & \multicolumn{1}{c}{SUMSS} & \multicolumn{1}{c}{NVSS} & \multicolumn{1}{c}{PMN} & \multicolumn{2}{c}{\textit{IRAS}} & \multicolumn{3}{c}{\textit{AKARI}} & \multicolumn{6}{c}{\textit{ISOPHOT}}\\
\textit{IRAS} & 843\,MHz & 1.4\,GHz & 4.85\,GHz & 3000\,GHz &  5000\,GHz & 2141\,GHz &  3330\,GHz &  4160\,GHz & 1470\,GHz &  1620\,GHz &  1860\,GHz &  2520\,GHz &  3150\,GHz &  4930\,GHz\\
 & (mJy) & (mJy) & (mJy) & (Jy) & (Jy) & (Jy) & (Jy) & (Jy) & (Jy) & (Jy) & (Jy) & (Jy) & (Jy) & (Jy)\\
\midrule
F00198-7926 & 36.7$\pm$3.0 & -- & -- & 2.9$\pm$0.3&3.1$\pm$0.2 & -- &2.4$\pm$0.2 & -- &-- &-- &-- &-- &-- &-- & \\
\hline
F00199-7426 & 40.8$\pm$3.6 & -- & -- & 6.4$\pm$0.5&4.2$\pm$0.3 & 5.4$\pm$0.5&4.4$\pm$0.3&3.9$\pm$0.3 & 3.0$\pm$0.9 & 4.2$\pm$1.3 & 5.3$\pm$1.6 & 7.5$\pm$2.3 & -- &-- &\\
\hline
F01268-5436 & 20.3$\pm$1.6 & -- & -- & 2.3$\pm$0.2&1.7$\pm$0.2 & 2.5$\pm$0.5 & 1.6$\pm$0.1 & -- &-- &-- &-- &-- &-- &-- & \\
\hline
F01388-4618 & 19.7$\pm$1.4 & -- & -- & 3.7$\pm$0.3&2.9$\pm$0.3 & -- &2.8$\pm$0.2 & -- &-- &-- &-- &-- &-- &-- & \\
\hline
F01419-6826 & 12.5$\pm$1.1 & -- & -- & 2.5$\pm$0.2&2.2$\pm$0.1 & -- &2.8$\pm$0.2 & -- &-- &-- &-- &-- &-- &-- & \\
\hline
F02364-4751 & 29.7$\pm$1.9 & -- & -- & 5.0$\pm$0.4&2.8$\pm$0.2 & 4.0$\pm$0.5 & 3.2$\pm$0.2 & -- &-- &-- &-- &-- &-- &-- & \\
\hline
F03068-5346 & 28.9$\pm$2.2 & -- & -- & 4.0$\pm$0.3&3.4$\pm$0.2 & -- &2.9$\pm$0.2 & 3.1$\pm$0.3 & 1.4$\pm$0.4 & 1.7$\pm$0.5 & -- &-- &-- &-- &\\
\hline
F03481-4012 & 21.9$\pm$1.7 & 15.8$\pm$1.1 & -- & 2.6$\pm$0.2&1.8$\pm$0.1 & -- &1.6$\pm$0.1 & -- &-- &-- &-- &-- &-- &-- & \\
\hline
F04063-3236 & 11.9$\pm$1.6 & 11.5$\pm$1.2 & -- & 2.1$\pm$0.2&1.8$\pm$0.1 & -- &1.7$\pm$0.2 & -- &-- &-- &-- &-- &-- &-- & \\
\hline
F06021-4509 & 10.9$\pm$1.3 & -- & -- & 2.4$\pm$0.3&1.6$\pm$0.1 & -- &1.4$\pm$0.1 & -- &-- &-- &-- &-- &-- &-- & \\
\hline
F06035-7102 & 152.0$\pm$10.7 & -- & -- & 5.7$\pm$0.4 & 5.1$\pm$0.3 & 4.8$\pm$0.3&4.5$\pm$0.2&4.6$\pm$0.3 & 1.4$\pm$0.4&2.0$\pm$0.6&2.9$\pm$0.9&5.2$\pm$1.6&5.0$\pm$1.5&5.9$\pm$1.8 & \\
\hline
F06206-6315 & 22.6$\pm$1.6 & -- & -- & 4.6$\pm$0.4&4.0$\pm$0.2 & 4.0$\pm$0.3&3.5$\pm$0.2&3.5$\pm$0.3 & 1.5$\pm$0.5&1.9$\pm$0.6&2.9$\pm$0.9&4.6$\pm$1.4&4.2$\pm$1.3&4.8$\pm$1.5 & \\
\hline
F18582-5558 & 12.0$\pm$1.2 & -- & -- & -- &1.9$\pm$0.2 & -- &1.8$\pm$0.1 & -- &-- &-- &-- &-- &-- &-- & \\
\hline
F20117-3249 & 54.4$\pm$3.5 & 57.1$\pm$3.4 & -- & -- &1.5$\pm$0.1 & -- &2.3$\pm$0.2 & -- &-- &-- &-- &-- &-- &-- & \\
\hline
F20445-6218 & 17.8$\pm$1.4 & -- & -- & 2.9$\pm$0.3&2.2$\pm$0.1 & 2.9$\pm$0.4 & 2.1$\pm$0.2 & -- &-- &-- &-- &-- &-- &-- & \\
\hline
F21178-6349 & 10.2$\pm$1.4 & -- & -- & 2.0$\pm$0.2&1.6$\pm$0.1 & -- &1.4$\pm$0.1 & -- &-- &-- &-- &-- &-- &-- & \\
\hline
F21292-4953 & 28.5$\pm$1.9 & -- & -- & 3.1$\pm$0.3&2.5$\pm$0.2 & 3.5$\pm$0.6 & 2.5$\pm$0.2 & -- &-- &-- &-- &-- &-- &-- & \\
\hline
F21295-4634 & 18.2$\pm$1.4 & -- & -- & 3.2$\pm$0.3&2.4$\pm$0.2 & -- &-- &-- & -- &-- &-- &-- &-- &-- & \\
\hline
F23389-6139 & 166.0$\pm$9.7 & -- & 59.0$\pm$8.5 & 4.3$\pm$0.3&3.6$\pm$0.2 & -- &-- &-- & 1.5$\pm$0.5&2.0$\pm$0.6&2.8$\pm$0.9&4.0$\pm$1.2&3.3$\pm$1.0&3.9$\pm$1.2 & \\
\hline
\end{tabular}
\end{table}

\tabcolsep=0.05cm
\begin{table}
\tiny{
\centering
\caption{An overview of all radio-continuum flux density measurements produced using ATCA data obtained under the project code C2993 as part of this study. `--' denote sources without a measurement for that central frequency. \label{table:atca_measurements}}
\begin{tabular}{lccccccccccccccccccccccccc}\toprule
 Source & \multicolumn{24}{c}{This Study} \\
 \textit{IRAS} & 1.5\,GHz & 1.6\,GHz & 1.8\,GHz & 2.0\,GHz & 2.2\,GHz & 2.4\,GHz & 2.6\,GHz & 2.8\,GHz & 4.4\,GHz & 4.5\,GHz & 5.0\,GHz & 5.5\,GHz & 5.6\,GHz & 6.3\,GHz & 6.8\,GHz & 7.2\,GHz & 7.3\,GHz & 8.3\,GHz & 8.8\,GHz & 9.2\,GHz & 17.0\,GHz & 21.0\,GHz & 43.1\,GHz & 48.1\,GHz\\
  & (mJy) & (mJy) & (mJy) & (mJy) & (mJy) & (mJy) & (mJy) & (mJy) & (mJy) & (mJy) & (mJy) & (mJy) & (mJy) & (mJy) & (mJy) & (mJy) & (mJy) & (mJy) & (mJy) & (mJy) & (mJy) & (mJy) & (mJy) & (mJy)\\
\midrule
F00198-7926 & 19.4$\pm$1.2 & -- & -- &14.9$\pm$1.0 & -- &12.3$\pm$0.9 & -- &11.0$\pm$0.9 &10.4$\pm$0.7 & -- &9.7$\pm$0.6 & -- &8.9$\pm$0.6 &8.0$\pm$0.6 & -- & -- &6.7$\pm$0.5 & -- &5.7$\pm$0.4 & -- &3.5$\pm$0.3 &2.7$\pm$0.3 &1.5$\pm$0.2 &1.4$\pm$0.3 &\\
\hline
F00199-7426 & 19.0$\pm$1.2 & -- & -- &16.9$\pm$1.1 & -- &14.3$\pm$0.9 & -- &13.3$\pm$0.9 &10.7$\pm$0.7 & -- &9.9$\pm$0.6 & -- &8.6$\pm$0.6 &7.8$\pm$0.6 & -- & -- &6.6$\pm$0.5 & -- &5.0$\pm$0.4 & -- &3.4$\pm$0.2 &2.5$\pm$0.3 &1.5$\pm$0.2 &2.0$\pm$0.3 &\\
\hline
F01268-5436 & 10.2$\pm$0.8 & -- & -- &8.4$\pm$0.7 & -- &7.4$\pm$0.7 & -- &7.2$\pm$0.6 &5.0$\pm$0.4 & -- &4.2$\pm$0.4 & -- &3.5$\pm$0.4 &3.4$\pm$0.3 & -- & -- &2.9$\pm$0.2 & -- &1.9$\pm$0.3 & -- &1.8$\pm$0.2 &1.4$\pm$0.3 &1.0$\pm$0.1 &1.1$\pm$0.2 &\\
\hline
F01388-4618 & 9.2$\pm$0.8 & -- & -- &8.3$\pm$0.7 & -- &8.4$\pm$0.7 & -- &7.8$\pm$0.6 &6.4$\pm$0.4 & -- &5.9$\pm$0.4 & -- &5.5$\pm$0.4 &4.6$\pm$0.3 & -- & -- &4.0$\pm$0.3 & -- &3.2$\pm$0.3 & -- &2.0$\pm$0.2 &1.8$\pm$0.3 &0.9$\pm$0.1 &1.2$\pm$0.2 &\\
\hline
F01419-6826 & 7.0$\pm$0.6 & -- & -- &7.0$\pm$0.5 & -- &6.6$\pm$0.6 & -- &4.8$\pm$0.7 &4.2$\pm$0.3 & -- &3.6$\pm$0.3 & -- &3.1$\pm$0.3 &2.9$\pm$0.2 & -- & -- &2.7$\pm$0.2 & -- &2.2$\pm$0.2 & -- &1.6$\pm$0.2 &1.0$\pm$0.2 & -- &1.3$\pm$0.3 &\\
\hline
F02364-4751 & 17.5$\pm$1.1 & -- & -- &14.0$\pm$0.9 & -- &12.1$\pm$0.8 & -- &12.2$\pm$0.8 &8.9$\pm$0.5 & -- &7.3$\pm$0.4 & -- &6.8$\pm$0.4 &5.8$\pm$0.4 & -- & -- &5.2$\pm$0.4 & -- &3.6$\pm$0.4 & -- &2.9$\pm$0.2 &2.2$\pm$0.2 &1.4$\pm$0.1 &1.2$\pm$0.2 &\\
\hline
F03068-5346 &  -- & -- &9.5$\pm$1.0 & -- & -- & -- &6.0$\pm$0.9 & -- &6.9$\pm$0.6 & -- &6.4$\pm$0.8 & -- &6.0$\pm$0.7 &5.4$\pm$0.6 & -- & -- &4.4$\pm$0.6 & -- &4.4$\pm$0.3 & -- &3.6$\pm$0.4 &2.7$\pm$0.3 &1.7$\pm$0.1 &2.0$\pm$0.2 &\\
\hline
F03481-4012 & 11.3$\pm$0.7 & -- & -- &10.5$\pm$0.6 & -- &8.7$\pm$0.6 & -- &8.3$\pm$0.6 &5.9$\pm$0.4 & -- &6.1$\pm$0.4 & -- &5.3$\pm$0.4 &4.8$\pm$0.3 & -- & -- &3.8$\pm$0.3 & -- &3.0$\pm$0.3 & -- &2.2$\pm$0.2 &1.8$\pm$0.3 &0.9$\pm$0.1 &0.7$\pm$0.2 &\\
\hline
F04063-3236 & 7.4$\pm$0.5 & -- & -- &6.5$\pm$0.5 & -- &5.5$\pm$0.5 & -- &5.3$\pm$0.5 &4.8$\pm$0.4 & -- &5.5$\pm$0.3 & -- &4.8$\pm$0.3 &5.0$\pm$0.3 & -- & -- &3.9$\pm$0.3 & -- &2.9$\pm$0.3 & -- &1.8$\pm$0.2 &1.3$\pm$0.3 &0.8$\pm$0.1 & -- &\\
\hline
F06021-4509 & 7.1$\pm$0.6 & -- & -- &7.5$\pm$0.6 & -- &6.8$\pm$0.6 & -- &5.8$\pm$0.6 &4.8$\pm$0.3 & -- &4.8$\pm$0.3 & -- &4.4$\pm$0.3 &4.2$\pm$0.2 & -- & -- &3.6$\pm$0.2 & -- &3.0$\pm$0.2 & -- &1.7$\pm$0.2 &1.1$\pm$0.2 &1.2$\pm$0.2 &1.3$\pm$0.3 &\\
\hline
F06035-7102 &  -- & -- & -- & -- & -- & -- & -- & -- &20.6$\pm$2.8 & -- &20.1$\pm$2.6 & -- &17.8$\pm$2.2 &16.1$\pm$2.0 & -- & -- &13.4$\pm$1.6 & -- &11.0$\pm$1.2 & -- &7.0$\pm$2.2 &5.1$\pm$2.3 &1.8$\pm$0.2 &2.0$\pm$0.3 &\\
\hline
F06206-6315 & 15.5$\pm$0.9 & -- & -- &14.6$\pm$0.9 & -- &14.7$\pm$0.9 & -- &13.2$\pm$0.8 &11.4$\pm$0.6 & -- &10.5$\pm$0.6 & -- &9.4$\pm$0.5 &8.8$\pm$0.5 & -- & -- &7.7$\pm$0.4 & -- &6.6$\pm$0.4 & -- &3.5$\pm$0.4 &2.8$\pm$0.3 &1.5$\pm$0.2 &2.3$\pm$0.3 &\\
\hline
F18582-5558 & 6.7$\pm$0.6 & -- & -- &7.1$\pm$0.6 & -- &7.0$\pm$0.6 & -- &7.4$\pm$0.6 &6.0$\pm$0.5 & -- &5.8$\pm$0.5 & -- &5.5$\pm$0.5 &5.0$\pm$0.5 & -- &4.8$\pm$0.5 & -- &4.3$\pm$0.4 & -- &3.9$\pm$0.3 &2.3$\pm$0.2 &2.0$\pm$0.3 &0.8$\pm$0.1 & -- &\\
\hline
F20117-3249 & 49.7$\pm$2.5 & -- & -- &44.7$\pm$2.3 & -- &41.2$\pm$2.2 & -- &36.8$\pm$2.0 &25.4$\pm$1.7 & -- &23.4$\pm$1.7 & -- &19.2$\pm$1.6 &16.2$\pm$1.4 & -- & -- &13.0$\pm$1.4 & -- &9.8$\pm$1.1 & -- &7.1$\pm$0.6 &5.1$\pm$0.6 &2.9$\pm$0.1 &2.3$\pm$0.2 &\\
\hline
F20445-6218 & 9.0$\pm$0.6 & -- & -- &8.2$\pm$0.6 & -- &7.1$\pm$0.5 & -- &6.0$\pm$0.5 &4.7$\pm$0.4 & -- &3.8$\pm$0.4 & -- &3.5$\pm$0.4 &3.2$\pm$0.3 & -- & -- &2.8$\pm$0.3 & -- &2.4$\pm$0.3 & -- &1.7$\pm$0.2 &1.4$\pm$0.2 &0.9$\pm$0.1 & -- &\\
\hline
F21178-6349 &  -- &5.1$\pm$0.5 & -- & -- &4.0$\pm$0.4 & -- & -- &2.8$\pm$0.5 & -- &2.2$\pm$0.2 & -- &1.6$\pm$0.3 & -- & -- &1.6$\pm$0.2 & -- & -- & -- &0.7$\pm$0.2 & -- &1.1$\pm$0.1 &0.8$\pm$0.3 & -- & -- &\\
\hline
F21292-4953 &  -- & -- &16.6$\pm$1.8 & -- & -- & -- &11.6$\pm$1.2 & -- & -- & -- &9.7$\pm$1.0 & -- & -- & -- &8.0$\pm$0.9 & -- & -- & -- &7.5$\pm$0.9 & -- &6.4$\pm$0.6 &4.4$\pm$0.7 & -- & -- &\\
\hline
F21295-4634 & 10.1$\pm$0.8 & -- & -- &10.2$\pm$0.8 & -- &7.5$\pm$0.8 & -- &4.4$\pm$0.7 &3.7$\pm$0.5 & -- &4.1$\pm$0.5 & -- &3.2$\pm$0.5 &3.0$\pm$0.5 & -- & -- &2.3$\pm$0.4 & -- &1.9$\pm$0.3 & -- &1.8$\pm$0.2 &1.1$\pm$0.3 &0.5$\pm$0.1 & -- &\\
\hline
F23389-6139 & 117.8$\pm$6.0 & -- & -- &100.4$\pm$5.1 & -- &91.3$\pm$4.7 & -- &82.3$\pm$4.3 &56.1$\pm$0.8 & -- &49.2$\pm$0.9 & -- &41.8$\pm$0.9 &37.1$\pm$0.9 & -- &28.2$\pm$1.0 & -- & -- &25.0$\pm$0.6 & -- &10.4$\pm$0.3 &8.3$\pm$0.6 &3.9$\pm$0.2 &3.8$\pm$0.3 &\\
\hline
\end{tabular}}
\end{table}
\end{landscape}

\section{Spectral Energy Distributions}
In this section we present all SEDs, with the most supported model overlaid, for each source. All available flux density measurements have been included.

\begin{minipage}{\linewidth}
\centering
\includegraphics[width=\linewidth]{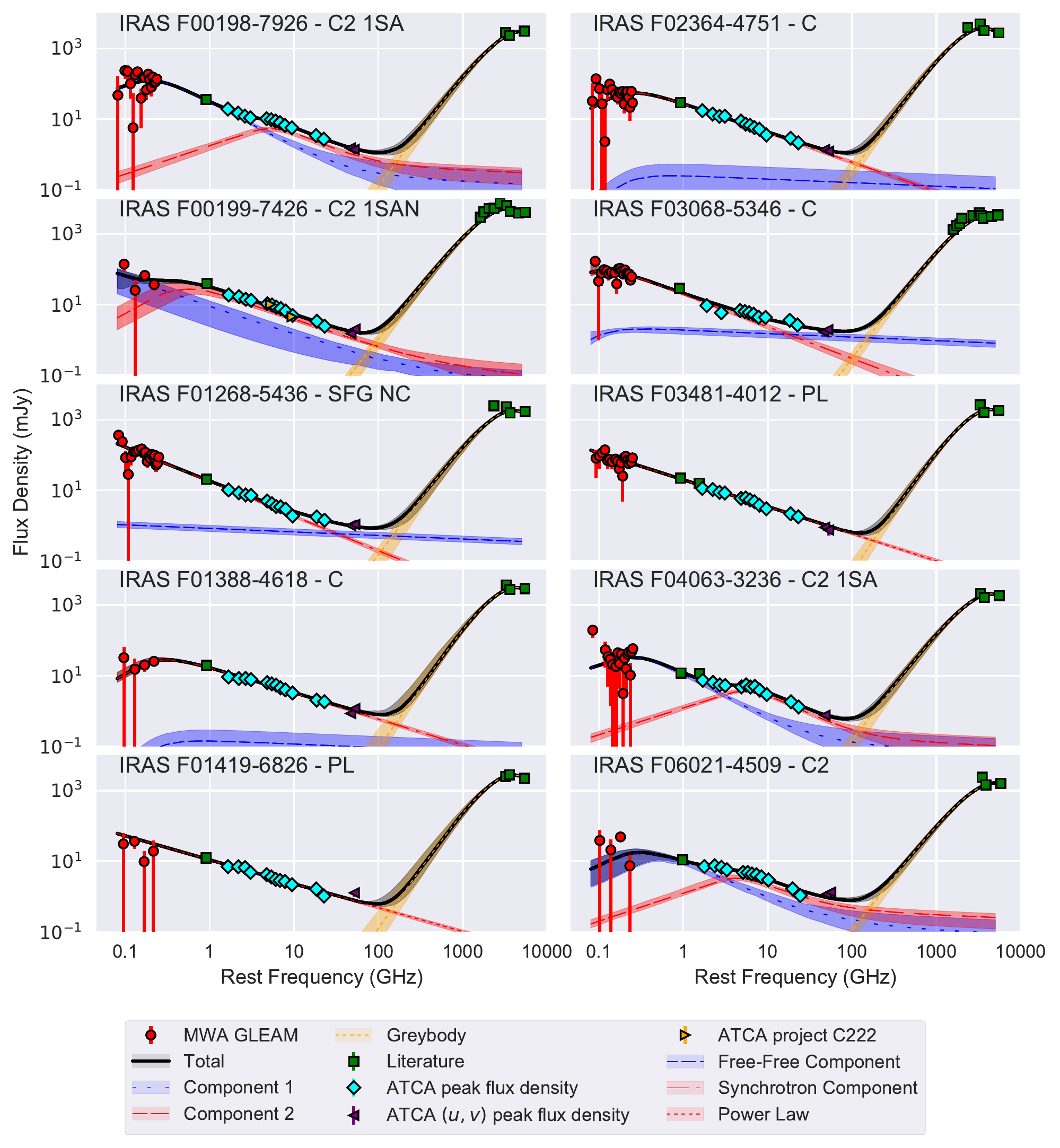}
\captionof{figure}{The observed data and preferred spectral energy distribution modelling of the SFGs from our sample in RA order. We include any components that make up the most preferred model focusing on the radio continuum. Highlighted regions represent the 1$\sigma$ uncertainty sampled by \emcee.  \label{fig:sed1}}
\end{minipage}

\begin{figure}
\ContinuedFloat
\centering
\includegraphics[width=\linewidth]{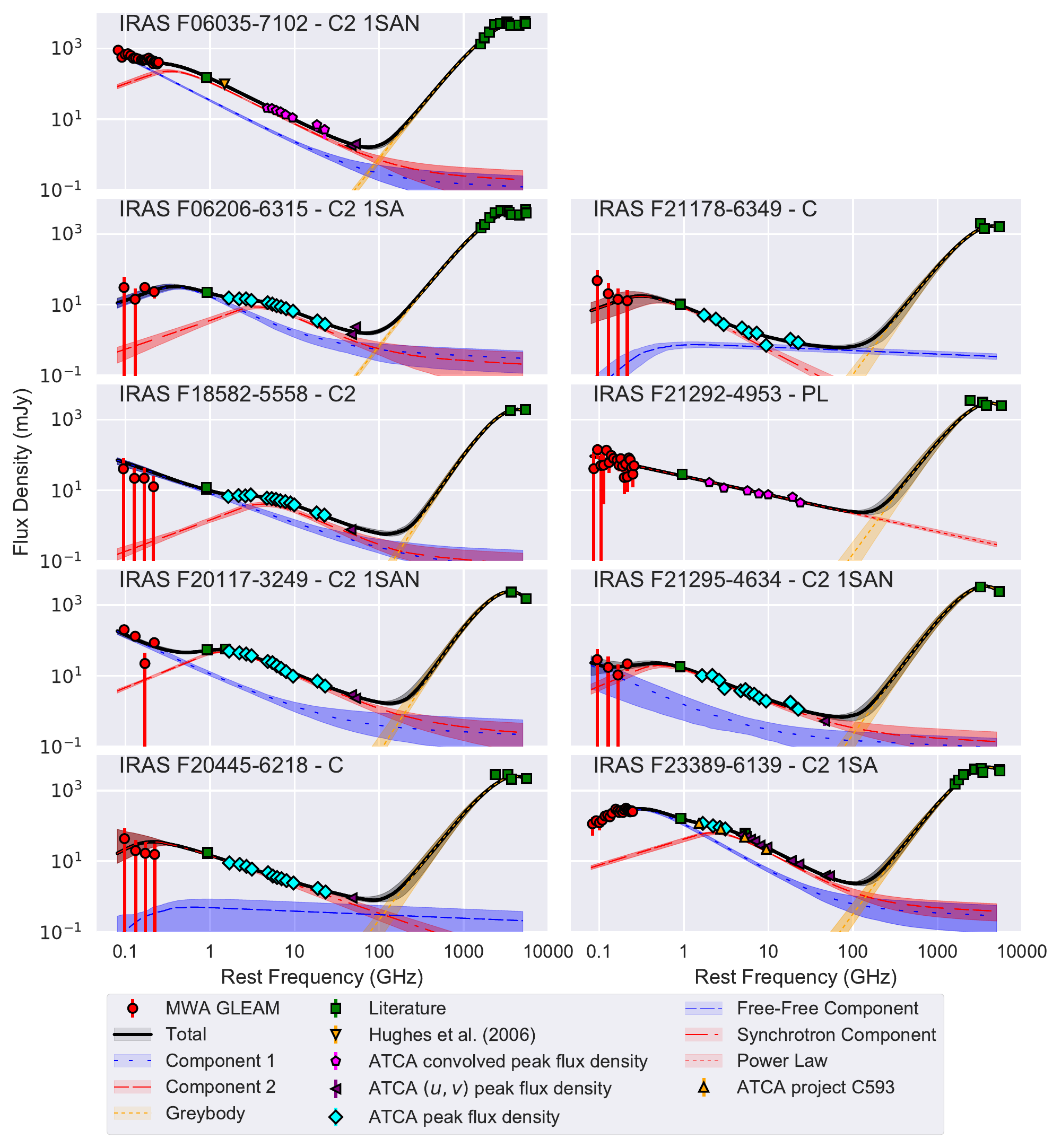}
\caption{\textit{(Continued)} The observed data and preferred spectral energy distribution modelling of the SFGs from our sample in RA order. We include any components that make up the most preferred model focusing on the radio continuum. Highlighted regions represent the 1$\sigma$ uncertainty sampled by \emcee.  \label{fig:sed2}}
\end{figure}

\end{document}